\documentclass[nonacm,sigconf]{acmart}
\usepackage{comment}
\usepackage{graphicx}
\usepackage{url}
\usepackage{listings}
\usepackage{amsmath}
\usepackage{amsthm}
\usepackage{xspace}
\usepackage{subcaption} 
\usepackage{caption,subcaption}
\usepackage{numberedblock}
\usepackage{adjustbox}
\usepackage{algorithm}  
\usepackage[noend]{algpseudocode}  
\usepackage{multirow} 

\usepackage{fancyhdr}

\usepackage{seqsplit}
\newcommand{\ttt}[1]{%
  \begingroup
    \protect\renewcommand{\seqinsert}{\ifmmode\allowbreak\else\-\fi}%
    \protect\texttt{\protect\seqinsert{\protect\seqsplit{\small#1}}}%
  \endgroup
}
\newcommand{\tttscript}[1]{%
  \begingroup
    \protect\renewcommand{\seqinsert}{\ifmmode\allowbreak\else\-\fi}%
    \protect\texttt{\protect\seqinsert{\protect\seqsplit{\scriptsize#1}}}%
  \endgroup
}
\newcommand{\tttfoot}[1]{%
  \begingroup
    \protect\renewcommand{\seqinsert}{\ifmmode\allowbreak\else\-\fi}%
    \protect\texttt{\protect\seqinsert{\protect\seqsplit{\footnotesize#1}}}%
  \endgroup
}

\usepackage{tikz}
\usetikzlibrary{backgrounds,fit, decorations.markings, patterns, shapes, positioning, arrows}
\usepackage{moeptikz}

\includecomment{techreport}

\newcommand{\update}[1]{#1}

\newcommand{\todo}[1]{}
\newcommand{\we}[1]{\textcolor{red}{[WE: #1]}}

\newcommand{\myparagraph}[1]{\vspace{0.25em}\noindent\textit{#1:}}

\newcommand{\system}{HoneyRoles\xspace}

\AtBeginDocument{%
  \providecommand\BibTeX{{%
    \normalfont B\kern-0.5em{\scshape i\kern-0.25em b}\kern-0.8em\TeX}}}

\setcopyright{acmcopyright}
\copyrightyear{2018}
\acmYear{2018}
\acmDOI{10.1145/1122445.1122456}

\begin{document}

\title{Role-Based Deception in Enterprise Networks}



\author{Iffat Anjum}
\affiliation{
  \institution{North Carolina State University}
  \city{Raleigh}
  \state{NC}
  \country{USA}
  }
\email{ianjum@ncsu.edu}

\author{Mu Zhu}
\affiliation{
  \institution{North Carolina State University}
  \city{Raleigh}
  \state{NC}
  \country{USA}
  }
\email{mzhu5@ncsu.edu}

\author{Isaac Polinsky}
\email{isaacpolinsky@ncsu.edu}
\affiliation{
  \institution{North Carolina State University}
  \city{Raleigh}
  \state{NC}
  \country{USA}
  }

\author{William Enck}
\email{whenck@ncsu.edu}
\affiliation{
  \institution{North Carolina State University}
  \city{Raleigh}
  \state{NC}
  \country{USA}
  }

\author{Michael K. Reiter}
\email{reiter@cs.unc.edu}
\affiliation{%
    \institution{University of North Carolina}
    \city{Chapel Hill}
    \state{NC}
    \country{USA}
    }

\author{Munindar Singh}
\email{mpsingh@ncsu.edu}
\affiliation{
  \institution{North Carolina State University}
  \city{Raleigh}
  \state{NC}
  \country{USA}
  }

\renewcommand{\shortauthors}{Anjum, et al.}
\begin{abstract}

Historically, enterprise network reconnaissance is an active process, often involving port scanning.
However, as routers and switches become more complex, they also become more susceptible to compromise.
From this vantage point, an attacker can passively identify high-value hosts such as the workstations of IT administrators, C-suite executives, and finance personnel.
The goal of this paper is to develop a technique to deceive and dissuade such adversaries.
We propose \system, which uses \emph{honey connections} to build metaphorical haystacks around the network traffic of client hosts belonging to high-value organizational roles.
The honey connections also act as network canaries to signal network compromise, thereby dissuading the adversary from acting on information observed in network flows.
We design a prototype implementation of \system using an OpenFlow SDN controller and evaluate its security using the PRISM probabilistic model checker.
\update{Our performance evaluation shows that \system has a small effect on network request completion time, and our security analysis demonstrates that once an alert is raised, \system can quickly identify the compromised switch with high probability.}
In doing so, we show that role-based network deception is a promising approach for defending against adversaries that have compromised network devices.

\end{abstract}


\keywords{Deception, Cyber Security, Software Defined Network, Data Plane}

\maketitle

\section{Introduction}

\begin{techreport}  
\todo{\begin{itemize}
  \item We need to reduce the emphasize on blackholing and SSL-striping.
	\item We need to make the statement of "importance of hiding the target or deception" more boost.
\end{itemize}}
Enterprises heavily rely on the security of their networks.
These networks often consist of a wide variety of computing resources, including desktops, laptops, servers, routers, and switches.
The resources support a range of activities by different types of users performing actions as different roles (e.g., IT administrators, C-suite executives, and finance personnel)~\cite{Casado:2009:Enterprise}.
By compromising one or more of these resources, an adversary may cause significant harm to the enterprise.
For example, it may steal credentials or access systems with the goal of exfiltrating sensitive information such as intellectual property and customer information, or modifying data such as source code repositories and payment systems.

The first phase of network infiltration is reconnaissance.
Traditional reconnaissance techniques such as port scanning are \emph{active}, and the current state-of-the-art network defenses have become highly tuned to identify them.
However, \emph{passive} reconnaissance by compromising packet forwarding devices and inspecting network flows to identify the existence and behaviors of client and server hosts is becoming increasingly feasible.
Specifically, as packet forwarding devices such as routers and switches become more complex, they become more prone to compromise~\cite{ye18,cim19,cox19}.
These targets include emerging Software Defined Networking (SDN) switches, which provide much broader and more flexible functionality~\cite{snowden:2014, Thimmaraju:Cloud:2016, Benton:2013:OVA, Markku:2014:Spook}.
\update{
  Prior solutions~\cite{MalRouter1, MalRouter2} seeking to defend against malicious forwarding devices are not directly applicable for SDN devices~\cite{SDN-RDC:2018}.
  Furthermore, SDN data plane defenses mostly concentrate on forwarding verification and other active attacks (e.g., packet delaying, tampering, dropping)~\cite{dhawan:2015:sphinx, VeriFlow:2013, shaghaghi:Wedgetail:2017,detectCompromisedSwitch:2015,Li:2018:Dyna}.
}

\update{
  The goal of this paper is to (1)~deceive adversaries by perturbing the network traffic information gained through passive reconnaissance, and (2)~dissuade an adversary from acting on observed information (e.g., performing active reconnaissance or an attack).
  }
We are particularly interested in protecting enterprise employees acting in high-value roles such as IT administrators, C-suite executives, and finance personnel.
Our vision is to build metaphorical ``haystacks'' around the network activities of these individuals.
The introduced network traffic perturbs reconnaissance, and if the adversary acts on the wrong intelligence, it will be detected with high probability, which will in effect dissuade the adversary from acting.

In this paper, we propose \system, which uses \emph{honey connections} to deceive adversaries using compromised packet forwarding devices for \emph{passive reconnaissance}.
\system coordinates honey connections by modeling fake hosts that are organized into roles corresponding to organizational functions of client hosts (e.g., IT administrator workstations).
\update{
\system performs integrity validation of honey connections such that they act as ``canaries'' for attacks against network clients.
}
In the event that an adversary modifies or blocks a honey connection, \system detects the adversary's existence and statistically identifies any compromised forwarding devices.

We evaluate the security of \system' defender-attacker environment using a probabilistic model checker (PRISM~\cite{PRISM}).
This simulation assumes an alert has been raised and measures the accuracy of detecting the location of compromised switches.
For a simulated Fat-Tree network topology with 50 real and 50 fake hosts and 1 compromised switch, we show that \system consistently ranks the compromised switch as most suspicious, regardless of the switch function (i.e., edge, aggregate, or core).
In the same environment with two compromised switches, we show that \system consistently ranks at least one of the switches as most suspicious.
The second compromised switch is also usually highly ranked, depending on its function.

We additionally used Mininet to emulate the Fat-Tree topology with 50 real and 50 fake hosts.
\update{When measuring the pairwise request completion time between real hosts and servers, we observed that \system has a small impact on network request completion time for a moderately loaded network (1 request per second per host on average). 
With a thorough experiment, we have seen that 90\% of hosts observe less than 14\% overhead in request completion time.}

This paper makes the following contributions:
\begin{itemize}

    \item \textit{We introduce role-based deception as an enterprise network defense.} 
    \system conceals the identity of critical client hosts and creates uncertainty for an adversary residing in one or more compromised packet forwarding devices.

    \item \update{\textit{We use honey connections to deflect and
	detect en route manipulation of client network traffic.} 
    \system uses statistical inference to identify any compromised network device.
}
    \item \update{\textit{We evaluate the security of \system's defender-attacker environment using a probabilistic model checker.}
    \system consistently tracks network events and successfully ranks the switches in terms of suspiciousness.}

\end{itemize}

The remainder of this paper proceeds as follows.
Section~\ref{sec:motivation} motivates our work.
Section~\ref{sec:overview} overviews \system's architecture and major goals.
Section~\ref{sec:design} describes the design principles.
Section~\ref{sec:security} provides a security analysis using a probabilistic model checker.
Section~\ref{sec:eval} evaluates performance overhead. 
Section~\ref{sec:discussion} discusses limitations. 
Section~\ref{sec:relwork} overviews related work. 
Section~\ref{sec:conc} concludes.

\end{techreport}
\section{Problem}
\label{sec:motivation}

\begin{techreport}
\todo{\begin{itemize}
	\item Depending on the updated comperative study (related work), we need to update our motivation statement.
	\item We need to reduce the emphasize on blackholing and SSL-striping.
	\item We need to make the statement of "importance of hiding the target or deception" more boost.
\end{itemize}}

Targeted attacks~\cite{Enbody:2013:Targeted,Li:2011:APT} and
threats to enterprise network infrastructure~\cite{SYNfulKnock:2015,CiscoSwitch:2018,Thimmaraju:SFHSF:2016} continue to increase.
Such attacks often begin with a foothold for reconnaissance.
Historically, footholds have been client workstations. 
However, network packet forwarding devices such as routers and SDN switches are becoming prime targets as they offer a valuable vantage point for reconnaissance and their increased complexity leaves them more prone to compromise. 

Once a foothold is established, the adversary performs reconnaissance
to identify targets that most profitably support its goals (e.g., to take
over the account of an IT administrator or C-suite executive).
\update{
From the vantage point of a compromised network switch, the adversary
can perform various en route network traffic attacks that strategically and selectively target high-value clients at critical times.
For example, it could inject malicious JavaScript into Web pages as they are returned from Web servers, or it could use SSL-stripping to eavesdrop on traffic and steal credentials.
%
Existing defenses such as HSTS have seen limited
deployment~\cite{Kranch:2015:UpgradingHTTPS}, in part because many developers do not understand how to use HSTS correctly, resulting in critical information such as login cookies being leaked.
 For networks that include mobile devices, Luo et al.~\cite{Luo:2019:TimeDoesNotHeal} found that popular mobile web browsers failed to fully support HSTS and were left open to clickjacking attacks.
Additionally, Krombholz et al.~\cite{Krombholz:2017:HTTPS} showed that TLS deployment is far too complex, leading to large numbers of incorrect HTTPS deployments.
%
Other attacks include redirecting client traffic to malicious servers or simply blackholing the traffic to keep a target from performing a critical task (e.g., monitoring IDS logs).
If done strategically and sparingly, such manipulation can fall under the detection thresholds of existing defenses~\cite{dhawan:2015:sphinx,detectCompromisedSwitch:2015,shaghaghi:Wedgetail:2017}.
}

Such attack activity can be broken down into three phases.
\emph{(1)~Passive reconnaissance:} the adversary passively intercepts
and tracks the communications of different organizational entities to identify the target roles' probable locations.
Other than forwarding collected data for further analysis, the adversary does not leave a trace for the defender to identify suspicious activity.
\update{
\emph{(2)~Active reconnaissance:} the adversary may perform a different type of active interception for pinpointing the target and increasing the confidence it has about the information.
Such activities may be detected by the defender; however, the adversary still does not disrupt communication.
\emph{(3)~Active attack:} the adversary has gained adequate confidence for target systems and decides to attack a client's network traffic.
Even if such activities raise an alarm, the adversary's location within the network may still be difficult to locate.
}

\update{
The three-phase attack plan described above demonstrates the danger of
reconnaissance as an important precursor to sophisticated attacks.
With information about the users, devices, and services on a network it is possible to design an attack strategy that minimizes the risk of detection.
For example, armed with information gathered passively, an adversary may realize its current foothold is unable to contact a sensitive server without triggering an alarm, resulting in it pivoting its foothold in the network to a device or user that can access the server.
For this reason, it is crucial to defend against network reconnaissance.
}

\myparagraph{Threat Model \& Assumptions}
\update{The goal of the adversary is to identify high value targets, learn enterprise secrets (e.g., intellectual property, customer data and credentials), and modify data en route to high-integrity servers (e.g., software code repositories, payment systems).  To do so, an adversary may target administrative systems, or connections to them, to gain access to target systems.  We assume the adversary is able to compromise one or more packet forwarding devices in the network.
From the vantage point of a forwarding device, the adversary can view, analyze, and modify all packets that flow through it.
We do assume that not \emph{all} of the forwarding devices are compromised, and that the defender can incrementally replace or refresh devices as they are detected.
}

We assume the adversary has some, but not all knowledge of the hosts in the network.
For example, we assume the IP addresses of important servers (e.g., Admin and Finance Servers) are known, based on other available information (e.g., DNS information).
However, we assume the adversary does not know the IP address and other details of workstations that perform specific organization roles (e.g., the IT Admin workstation).
Additionally, by compromising an SDN switch, the adversary has access to the SDN southbound network and hence can attempt to forge forwarding rules (e.g., OpenFlow messages) in the corresponding switch.
Finally, in order to achieve its goals, the adversary seeks to remain \emph{undetected}.
For example, if the defender identifies \emph{which} SDN switch is compromised, it will replace the switch, and the adversary will lose its foothold.

Our trusted computing base (TCB) includes the system defender (SDN controller or a separate trusted server) and the southbound network between the SDN controller and SDN switches.
As such, we assume the SDN switches are configured to either use out-of-band communication, or in-band communication protected by TLS.
We do not blindly assume that SDN switches are trustworthy.
Our security analysis in Section~\ref{sec:security} evaluates scenarios with one or two compromised switches.
Similarly, \system trusts its host agents running on workstations and servers.
\update{Finally, we assume the topology has a sufficient number of redundant forwarding paths for the ease of dynamic path management, discussed in Section~\ref{sec:design}. }

\end{techreport}

\section{Overview}\label{sec:overview}

\begin{techreport}
\todo{\begin{itemize}
	\item Clarify the, we are not only replaying the traced packet (we are creating actual session). 
	\item Articulate the fact that HoneyRole can work with other protocols as long as HTTP/HTTPS.
\end{itemize}}

\system seeks to use deception to mitigate the threat of compromised packet forwarding devices (e.g., switches and routers).
\update{
From the vantage point of a packet forwarding device, an adversary can
perform passive reconnaissance to identify high-value client hosts
(e.g., IT administrators, C-suite executives, and finance personnel),
active reconnaissance (e.g., selective probing or rerouting), and
perform en route traffic attacks (e.g., injecting content, SSL-stripping, blackholing).}
Our vision is to introduce honey network traffic that (1) deceives the adversary by building metaphorical ``haystacks'' around the network activities of high-value client hosts, 
and (2) dissuades the adversary from acting on information in real network traffic for fear of being detected.

Achieving this vision requires overcoming the following research challenges:

\begin{enumerate}

  \item[\textbf{C1}] \textbf{(Detection):} \textit{Compromised packet forwarding devices are difficult to detect.}
    An adversary performing passive monitoring will not produce detectable
    actions until it attempts an attack (possibly months after compromise), at
    which point it may be too late to detect a compromise.

  \item[\textbf{C2}] \textbf{(Exposure):} \textit{The adversary may have knowledge of some enterprise network components.}
    Information from DNS and publicly accessible websites~\cite{amass:2019, HT:2019} make hiding the identity of servers futile. \update{Servers receive a disproportionate amount of inbound connections, allowing an in-network adversary to distinguish between clients and servers, which may limit the effectiveness of moving-target defenses~\cite{decoy2018,decoyIP}.}

  \item[\textbf{C3}] \textbf{(Visibility):} \textit{The adversary may be aware of the deception system.}
    Na\"ively sending honey traffic is not effective if the adversary is aware of the defense.
    External events (e.g., stock market changes and DDoS attacks) can cause certain high-value client hosts to act predictably.

\end{enumerate} 

\begin{figure}[t]
  \centering
  \newcommand*{\nodelabel}[1]{{\small\bfseries\ttfamily #1}}
\scalebox{0.75}{
\begin{tikzpicture}[
    myclient/.style={client,minimum size=8mm},
    myswitch/.style={switch,fill=white,minimum size=8mm},
    myserver/.style={server,minimum size=8mm},
    link/.style={black,line width=1pt},
    clink/.style={black,dashed,line width=0.5pt},
    mycloud/.style={cloud,cloud puffs=10,draw=black,line width=1pt},
    role1/.style={circle,draw=black,fill=black!30,inner sep=0pt,minimum width=2mm,minimum height=2mm,line width=0.5pt},
    role2/.style={diamond,draw=black,fill=black!30,inner sep=0pt,minimum width=2mm,minimum height=2mm,line width=0.5pt},
    role3/.style={rectangle,draw=black,fill=black!30,inner sep=0pt,minimum width=2mm,minimum height=2mm,line width=0.5pt},
    role1fake/.style={circle,draw=black,fill=black!80,inner sep=0pt,minimum width=2mm,minimum height=2mm,line width=0.5pt},
    role2fake/.style={diamond,draw=black,fill=black!80,inner sep=0pt,minimum width=2mm,minimum height=2mm,line width=0.5pt},
    role3fake/.style={rectangle,draw=black,fill=black!80,inner sep=0pt,minimum width=2mm,minimum height=2mm,line width=0.5pt},
    mytext/.style={draw=none,fill=none,right}
]

\node[myclient,label=left:\nodelabel{$c_1$}] (c1) at (0,3) {};
\node[myclient,label=left:\nodelabel{$c_2$}] (c2) at (0,2) {};
\node[myclient,label=left:\nodelabel{$c_3$}] (c3) at (0,1) {};
\node[myclient,label=left:\nodelabel{$c_4$}] (c4) at (0,0) {};

\node[myserver,label=right:\nodelabel{$s_1$}] (s1) at (8,3) {};
\node[myserver,label=right:\nodelabel{$s_2$}] (s2) at (8,2) {};
\node[myserver,label=right:\nodelabel{$s_3$}] (s3) at (8,1) {};
\node[myserver,label=right:\nodelabel{$s_4$}] (s4) at (8,0) {};

\node[myswitch,label=:\nodelabel{$sw1$}] (sw1) at (1.5,2.5) {};
\node[myswitch,label=below:\nodelabel{$sw_2$}] (sw2) at (1.5,0.5) {};

\node[myswitch,label=:\nodelabel{$sw_7$}] (sw7) at (6.5,2.5) {};
\node[myswitch,label=below:\nodelabel{$sw_8$}] (sw8) at (6.5,0.5) {};

\draw[link] (c1) -- (sw1);
\draw[link] (c2) -- (sw1);
\draw[link] (c3) -- (sw2);
\draw[link] (c4) -- (sw2);

\draw[link] (s1) -- (sw7);
\draw[link] (s2) -- (sw7);
\draw[link] (s3) -- (sw8);
\draw[link] (s4) -- (sw8);

\node[mycloud,minimum width=4.2cm,minimum height=3.5cm,label=below:\nodelabel{Core Network}] (core) at (4,1.5) {};
\node[myswitch,label=:\nodelabel{$sw_3$}] (sw3) at (3.25,2) {};
\node[myswitch,label=below:\nodelabel{$sw_4$}] (sw4) at (3.25,1) {};
\node[myswitch,fill=red,label=:\nodelabel{$sw_5$}] (sw5) at (4.75,2) {};
\node[myswitch,label=below:\nodelabel{$sw_6$}] (sw6) at (4.75,1) {};

\draw[link] (sw1) -- (sw3);
\draw[link] (sw1) -- (sw4);
\draw[link] (sw2) -- (sw3);
\draw[link] (sw2) -- (sw4);

\draw[link] (sw7) -- (sw5);
\draw[link] (sw7) -- (sw6);
\draw[link] (sw8) -- (sw5);
\draw[link] (sw8) -- (sw6);

\draw[link] (sw3) -- (sw4);
\draw[link] (sw3) -- (sw5);
\draw[link] (sw3) -- (sw6);

\draw[link] (sw4) -- (sw5);
\draw[link] (sw4) -- (sw6);

\draw[link] (sw5) -- (sw6);

\node[myserver,label=above:\nodelabel{SDN Controller}] (controller) at (4,4.0) {};
\draw[clink] (controller) -- (sw1);
\draw[clink] (controller) -- (sw2);
\draw[clink] (controller) -- (sw7);
\draw[clink] (controller) -- (sw8);
\draw[clink] (controller) -- (core);

\node[mytext] (real) at (-1,-1) {\nodelabel{Real:}};
\node[role1,label=right:\nodelabel{Role 1}] (r1r) at (0.25,-1.0) {};
\node[role2,label=right:\nodelabel{Role 2}] (r2r) at (2.0,-1.0) {};
\node[role3,label=right:\nodelabel{Role 3}] (r3r) at (3.75,-1.0) {};
\node[mytext] (fake) at (-1,-1.5) {\nodelabel{Fake:}};
\node[role1fake,label=right:\nodelabel{Role 1}] (r1f) at (0.25,-1.5) {};
\node[role2fake,label=right:\nodelabel{Role 2}] (r2f) at (2.0,-1.5) {};
\node[role3fake,label=right:\nodelabel{Role 3}] (r3f) at (3.75,-1.5) {};

\node[role1,yshift=7pt,right=2pt of c1] (c1r1) {};
\node[role2fake,yshift=0pt,right=2pt of c1] (c1f2) {};
\node[role3fake,yshift=-7pt,right=2pt of c1] (c1f3) {};

\node[role1fake,yshift=7pt,right=2pt of c2] (c2f1) {};
\node[role2,yshift=0pt,right=2pt of c2] (c2r2) {};
\node[role3fake,yshift=-7pt,right=2pt of c2] (c2f3) {};

\node[role1fake,yshift=7pt,right=2pt of c3] (c3f1) {};
\node[role2fake,yshift=0pt,right=2pt of c3] (c3f2) {};
\node[role3,yshift=-7pt,right=2pt of c3] (c3r3) {};

\node[role1fake,yshift=7pt,right=2pt of c4] (c4f1) {};
\node[role2fake,yshift=0pt,right=2pt of c4] (c4f2) {};
\node[role3fake,yshift=-7pt,right=2pt of c4] (c4f3) {};


\draw[blue,line width=1.5pt] plot [smooth] coordinates {(c1r1) (sw1) (sw3) (sw6) (sw7) (s1)};
\draw[red,line width=1.5pt] plot [smooth] coordinates {(c3f1) (sw2) (sw4) (sw5) (sw7) (s1)};

\end{tikzpicture}
}
  \caption{Overview of \system}
  \label{fig:overview}
\end{figure}

Figure~\ref{fig:overview} overviews the high-level intuition behind \system.
The figure depicts four client hosts ($c_1$-$c_4$) and four servers ($s_1$-$s_4$) connected by a network topology with redundant links and switches.
A network administrator partitions each client host based on \emph{organizational roles}.
In the figure, $c_1$ is in \texttt{Role 1}, $c_2$ is in \texttt{Role 2}, $c_3$ is in \texttt{Role 3}.
Host $c_4$ is not assigned to any role.
\system then installs software agents (honey agents) on client hosts that produce fake network traffic.
Each client host is assigned at least one honey agent, each of which is assigned one of the organizational roles.
Some physical hosts may run multiple honey agents.

\update{
\system uses hosts with honey agents to establish \emph{honey connections} with the real servers (e.g., SMTP, HTTP, FTP).
Using \emph{honey connections}, honey agents establish \emph{new}
application-layer protocol sessions with the servers (simply replaying
network traces would be detectable). 
}
\system uses Software-Defined Networking (SDN) to dynamically change the forwarding paths between client hosts (both honey and real) and servers.
Each forwarding path is selected randomly and applies to both real and fake (honey) connections.
Figure~\ref{fig:overview} shows two forwarding paths.
The blue path ($c_1$-$s_1$) containing real traffic avoids the compromised switch ($sw_5$), and the red path ($c_3$-$s_1$) containing fake (honey) traffic passes through the compromised switch ($sw_5$).
Note that the blue path could have just as easily passed through the compromised switch.
The goal of deception is to provide the adversary with sufficient fake information such that it does not know what information to believe.
Furthermore, \system's honey connections act as canaries.
\update{
If the adversary guesses wrong and performs an en route traffic attack on a honey connection, \system will detect the adversary's existence.}
If the adversary guesses wrong enough times, \system statistically identifies the compromised forwarding device.

\system addresses the \emph{Detection} challenge by increasing the frequency at which a given compromised packet forwarding device will see network traffic that may be viewed by the adversary as ``valuable.''
For example, a switch near client host $c_4$ (Figure~\ref{fig:overview}) may not normally see network traffic to a domain controller. 
However, honey connections from a honey agent on $c_4$ provide the delusion that domain controller traffic can go through that switch.
In fact, this design choice enables \system to  not merely overcome
but embrace the
\emph{Exposure} challenge.
By not obfuscating the network identities of high-value servers, \system uses honey connections to them as bait.

\system addresses the \emph{Visibility} challenge through its use of organizational roles to parameterize the creation of honey connections.
For each role, an administrator specifies a \emph{role profile}.
The profile defines: 
(1)~the number of real hosts,
(2)~the identities of the real hosts,
(3)~the number of honey agents,
(4)~the locations and identities for the honey host agents, and
(5)~the set of target servers~$S$ that are relevant to the role (e.g., domain controller and HR server).
\system assumes the adversary would attempt to identify target client hosts based on their connections to the high-value servers.
Therefore, \system monitors the network activity between real clients and high-value servers.
It uses these traffic patterns to automatically configure the honey host agents to send honey connections to the target servers with similar request rates.
Assume there are $r$ real client and $h$ honey host agents in same target role. 
Hence, the adversary has a $\frac{r}{r + h}$ chance of correctly identifying a real client host.
Note that acting on a wrong guess has negative consequences for the adversary (i.e., detection).

Importantly, \system does not provide any signal on detection (no change of strategy) toward the adversary, unless network administrators reconfigure a compromised switch.
Section~\ref{sec:security} evaluates the resulting defender-attacker outcomes using the PRISM probabilistic model checker. 

\update{
Finally, to ensure the adversary cannot distinguish a honey host from a real host, \system assumes an ambient network traffic generator to represent the general activity that a given user may perform~\cite{Swing:Vishwanath:2009, Sommers:2004:SNT, Bowen:2012:SGI}.
We note that these prior works are simply examples.
Designing and evaluating network traffic generators that can evade detection from modern machine learning algorithms is an orthogonal challenge.
\system would directly benefit from any advancements in this area.
However, we note that in our scenario, the classification must be performed on a compromised switch, which limits the computational capacity of the adversary, thereby limiting which machine learning algorithms can be used.
}
\end{techreport}
\section{Design}
\label{sec:design}

\begin{techreport}

This section discusses three considerations in the design of \system:
(1) \emph{honey connections}, including how they are managed by the software agents and how they are coordinated by the \system controller;
(2) dynamic \emph{forwarding path management} to distribute honey connections across many potentially compromised switches and help statistically identify compromised switches; and
(3) the \emph{belief maintenance system} within the \system controller, which is used to rank switches based on their probability of being compromised.

\subsection{Honey Connections}
\label{sec:connections}
Honey connections provide the primary form of deception in \system.
For expository reasons, we describe how to create honey connections
for a single role; it is straightforward to extend the design to an
arbitrary number of roles.

\subsubsection{Role Profile}
\label{sec:profile}
For each role, the administrator defines a fixed set $ID_r$ of real hosts matching their organizational roles.
The administrator defines for each role a honey host factor, $\alpha
\ge 0$, which yields the size of the set $ID_h$ of honey hosts for the
specific role. Specifically, $|ID_h| = \lceil \alpha \cdot |ID_r| \rceil$.
We expect a typical deployment will include at least as many honey hosts as real hosts ($\alpha \ge 1$).

\system identifies each host (both real and honey) via a 5-tuple:
$\mathtt{<ip, mac, type, role, switch>}$, where
$\mathtt{ip}$ is the host's IP address,
$\mathtt{mac}$ is the host's MAC address,
$\mathtt{type}$ indicates if the host is real or honey,
$\mathtt{role}$ specifies the host's organizational role (e.g., IT administrator), and
$\mathtt{switch}$ specifies the network switch to which the host is attached.
The $\mathtt{ip}$, $\mathtt{mac}$, and $\mathtt{switch}$ are fixed for
real hosts and are randomly assigned by \system for honey hosts.
Of these values, the $\mathtt{switch}$ has the most impact on the utility of honey connections, as it determines where in the network the honey host exists, and hence the other switches that honey connections to or from this switch will likely traverse.
\update{The honey agent (host) composition and assignment is further discussed in Section~\ref{subsubsec:honeyagent}.}

Finally, the role profile contains a set of target servers $S$, each associated with the organizational role.
Conceptually, $S$ defines the set of servers that users in a role connect with to perform their duties.
For example, for an IT administrator role, $S$ may include a domain controller, a centralized VM management server, and a configuration management system server.
For each server $s \in S$, the network administrator specifies information for valid connections (e.g., an unprivileged user and associated credentials) so that the size and frequency of packets in TLS-protected connections are indistinguishable from the connections generated by real hosts.

\subsubsection{Honey Agent}\label{subsubsec:honeyagent}
We envision that honey agents will reside on the same physical hardware as real hosts to reduce capital expenditures required for deploying \system.
However, network administrators can deploy hosts without users, e.g., decommissioned
computers, that run only honey agents.

The honey agent needs to be a privileged process capable of using distinct IP and MAC addresses.
This is achievable using either operating system virtualization or containerized environments.
For example, Qubes OS uses a hypervisor to containerize multiple distinct execution environments.
It provides a flexible and modular networking environment that can bridge virtual interfaces to different environments.
Alternatively, for non-hypervisor hosts, the honey agent could be deployed as a container.
For example, Docker can use bridge mode to give a container its own IP address.

Since our performance evaluation in Section~\ref{sec:eval} uses Mininet~\cite{Mininet}, we emulate the existence of a honey agent by creating extra Mininet hosts (with individual IP addresses) tagged as honey hosts. 
These extra honey hosts perform the necessary logic to act as honey agents, generating traffic according to assigned role and traffic distribution. 

\todo{ Articulate the fact that HoneyRole can work with other protocols as long as HTTP/HTTPS.}
\todo{Clarify the, we are not only replaying the traced packet (we are creating actual session)}

\subsubsection{Honey Agent Coordination}
\label{sec:heartbeat}
The \system controller coordinates the honey connections sent by honey agents using TLS-protected heartbeat messages.
It is important that heartbeats are sent on regular intervals and are statistically similar in size, as they are sent through the data plane and are observable by adversaries.
To preserve the size,
the controller can split the necessary information across heartbeats
or pad a heartbeat to achieve the correct size. Heartbeat messages are sent to real hosts to prevent the adversary from using heartbeats to identify honey hosts.

\update{
The purpose of the heartbeat messages is to parameterize the creation of honey connections.
To that end, each heartbeat contains the following information:
(1)~destination information (MAC address, IP address, transport-layer port),
(2)~number of RREs (Request Response Exchange),
(3)~RRE interval,
(4)~application-layer protocol information, and
(5)~estimated timeout.
The generation of believable honey connections additionally requires realistic application-layer content or information.
The application-layer protocol information depends on the type of protocol (e.g., SMTP, IMAP, FTP, HTTP). 
For example, HTTP/HTTPS connections may require a URL, cookies, and username/password pairs.  
Other application-level information can be Gmail cookies, protocol payloads (i.e., email bodies), passwords for unencrypted protocols (e.g., SMTP, POP, IMAP).
For simplicity, our implementation considers only HTTP and HTTPS traffic. 
}

\todo{We have search for some recent papers that can address all the traffic generation concerns (R1). Most importantly we have to how defend the argument in some ingenious way.}

\myparagraph{Capturing Real Traffic Profiles}
As described in Section~\ref{sec:overview}, a key idea of \system is that honey connections for a given role follow the traffic patterns of that role.
Existing traffic tracing and monitoring tools~\cite{Fonseca:2007:XPN, FlowMon, Wireshark} use multiple network sensors distributed throughout a network.
We achieve a similar capability using OpenFlow's flow-level statistics collection mechanism~\cite{OpenNetMon:2014:Adrichem, OpenSample:2014:Suh, PayLess:2014:Chowdhury}. 
Our implementation leverages this information within the ONOS SDN controller.
We leverage the OpenFlow control messages (e.g., PacketIn, FlowMod, FlowRemoved, FlowStatistics) to capture the near-realtime traces of real host connections. 

\myparagraph{Replicating Real Traffic Profiles}
Our implementation does not include ambient network traffic, but focuses on dynamically turning captured real traffic profiles into honey connections.
The role profile (Section~\ref{sec:profile}) defines a set of target servers $S$ that are relevant to the tasks of a given role.
\system generates honey connections of a specific role by observing the network connections between the real hosts $ID_r$ of that role and the corresponding servers in $S$.
As in Harpoon~\cite{Sommers:2004:SNT}, \system parameterizes traffic generation based on the following information for each time interval:
(1)~the \emph{source and destination addresses};
(2)~the \emph{payload size} for each source-destination pair;
(3)~the average \emph{number of active sessions} between each source-destination pair;
(4)~the \emph{time duration} based on an empirical distribution of time between consecutive connections as well as the average inter-arrival time;  and
(5)~\emph{header information} based on the distribution of common values such as MAC address, protocol, and port.

\todo{Reduce emphasize on blackholing and SSL-striping.}
\subsubsection{Honey Agent Reports}
\label{sec:report}
A honey agent sends reports as heartbeat responses.
Note that real hosts must also send reports (though without meaningful content)
to make them indistinguishable from honey hosts.
At a high level, a honey agent report provides a status update on the honey connections specified in previous heartbeats.
\update{Each alert included in a report specifies:
(1)~total number of requests sent, and
(2)~alert details (e.g., average delay, number of dropped request).}
Section~\ref{sec:algorithm} discusses how \system uses reports to statistically identify the locations of compromised switches.
\update{
Recall that honey connections act as canaries for the existence of a network adversary.
We envision a \system deployment will include a collection of detection types (e.g., packet rerouting, packet hijacking, manipulation) for different types of applications (e.g., SMTP, FTP).
}

Our implementation detects two attack types: SSL-stripping and blackholing.
SSL-stripping occurs when the victim first visits the HTTP version of a website.
Normally, the server will redirect the web browser to the HTTPS version of the website.
However, an en route network adversary can suppress the redirection to keep the victim using HTTP URLs,
potentially revealing passwords or other security-sensitive information.
To detect SSL-stripping, \system uses honey connections that simulate the user entering just the domain name into the URL bar of the web browser.
If the honey agent does not receive the expected redirect to the HTTPS version of the web page, an alert is reported.

Network blackholing occurs when an in-network adversary prevents packets from reaching their destination.
For example, an adversary may wish to prevent an IT administrator from accessing a network logging server while it is performing an attack.
To detect blackholing, \system simply sends honey connections to the important target servers.
If a connection exceeds a pre-specified timeout period, an alert is reported.
However, normal network congestion and load at the target server can also cause honey connections to time-out.
Therefore, the belief maintenance system (Section~\ref{sec:algorithm}) must take care when using alerts of this type.

\update{We envision that detection algorithms for other attacks can be integrated in \system, e.g., SSL downgrade, wrong SSL certificate, page contents modified.}
However, the SSL-stripping and blackholing detectors are sufficient to demonstrate how heartbeats and reports function, \update{because they cover the spectrum of modification and dropping.}

\subsection{Forwarding Path Management}
\label{sec:paths}

\system dynamically changes the forwarding path from clients to servers to distribute honey connections across potentially compromised switches.
The dynamic forwarding path helps identify the location of a compromised switch.
Since the goal of the adversary is to distinguish between real and honey connections, it is important to minimize the differences between them.
Therefore, \system does not differentiate real connections from honey connections when changing forwarding paths.

A dynamic forwarding path helps identify the existence of an adversary more quickly, simply because it distributes packets in honey connections across more switches.
To better understand how the dynamically forwarding path helps identify the location of a compromised switch, consider a collection of alarms raised for honey connections between client $c_1$ and server $s_1$.
If the honey connections always traverse the same set of network switches, it is difficult to determine which switch is compromised.
However, if the forwarding path differs for each alarm, the
intersection of the forwarding paths can be used to isolate a compromised switch.
The belief maintenance system in Section~\ref{sec:algorithm} uses this intuition, but considers alarms in the network as a whole.

\system builds upon the OpenFlow SDN protocol to perform dynamic forwarding paths.
A key component of all SDN controllers (e.g., ONOS) is a reactive forwarding path algorithm that determines the best path from a source to a destination.
Network topologies commonly have redundant links and switches (e.g., see Figure~\ref{fig:overview}), and this forwarding path algorithm must avoid forwarding loops and potentially react to network congestion.

\update{We observe that given a network topology with sufficient redundancy, there will be multiple optimal paths within each pair of source and destination.}
Furthermore, the dynamic forwarding path can include slightly non-optimal forwarding paths as acceptable.
\system randomly selects from the set of acceptable forwarding paths.

\newcommand{\smac}{\mathtt{s\_mac}}
\newcommand{\dmac}{\mathtt{d\_mac}}
\newcommand{\sip}{\mathtt{s\_ip}}
\newcommand{\dip}{\mathtt{d\_ip}}
\newcommand{\dport}{\mathtt{d\_port}}

\update{More specifically, \system's dynamic path selection operates as follows.
  \system defines network flows as a 5-tuple: source IP address ($s_{ip}$), source transport-layer port ($s_{port}$), destination IP address ($d_{ip}$), destination transport-layer port ($d_{port}$), and transport-layer $protocol$ (i.e., TCP or UDP).
Whenever a new connection (honey or real) is set up by a source-destination pair, a PacketIn message (request for setting up a forwarding path) is sent to the controller by the edge switch connected with the source host.
\system's reactive forwarding application determines a maximal set of disjoint paths.}
Depending on the system requirement, this application can consider optimal disjoint paths only, or both optimal and non-optimal disjoint paths, or tolerate a certain percentage of overlap.
From the set of possible forwarding paths, \system selects a path using uniform random distribution.
Even if the defender suspects compromised switches on a certain path, it should not set a priority in the selection process, as this may be detected by the adversary, thereby revealing some of the defender's knowledge.

Given a topology with $p$ disjoint paths (both optimal and non-optimal), the probability of selecting a certain path is $1/p$.
At a given time $t$, there are $r$ real and $h$ honey connections for a given target server.
If there is a compromised switch in only one disjoint path, the probability that the adversary will be able to scan a real connection is $\frac{r}{p(r+h)}$.
Consequently, combining the dynamic forwarding and honey connections, \system builds a dense haystack around the real connections, making passive reconnaissance harder.

\subsection{Belief Maintenance System}
\label{sec:algorithm}

The goal of the belief maintenance system (BMS) is to alert the system administrator about the existence of an adversary, as well as potential locations of compromised switches.
However, it does not seek to precisely determine a specific switch or set of switches that are compromised.
Instead, the BMS ranks switches based on a level of suspiciousness.
The goal is to ensure all compromised switches are among the most
suspicious ones in the ranked list.  The BMS can reside on the SDN controller or on a separate server.

As discussed in Section~\ref{sec:report}, detection of adversarial activity and alert generation is performed by the honey agents.
Recall that \system uses both role-based honey connections and dynamic forwarding paths to entice the adversary into acting on false information.
\system cannot be certain about the network's adversarial state.
For example, some alarms (e.g., packet dropping) can be generated from either network failure or adversarial activity.
Furthermore, even for true positives for a given forwarding path
with $n$ switches, there is only a $\frac{1}{n}$ chance that a given switch is the source of the alarm. 
Therefore, the BMS maintains an updated mapping between the honey
connections and the corresponding forwarding paths and uses alarms from honey agent reports to update its belief of suspiciousness for each switch.

The BMS updates its current belief for each switch after each discrete time interval $\gamma$.
That is, if the current time is $t$, the next update will occur at $t+\gamma$.
The BMS uses the $\gamma$ period to collect statistics for the interval, after which the reports can be discarded.
For each switch $s_k$, the BMS calculates following 
for the time interval.
\begin{itemize}
	\item $a_{k}$: \update{number of alarms received for forwarding paths that include switch $s_k$ }
    \item $c_{k}$: number of honey connections forwarded by $s_k$ 
\end{itemize}
The BMS then calculates a risk factor $r_{k,t} = \frac{a_{k}}{c_{k}}$ for switch $k$ on a specific time $t$.
It computes an overall risk factor $R_{k,t}$ for switch $s_k$ 
using exponential moving average (where $R_{k,0} = r_{k,0}$):
\begin{equation} \label{eq:crateUpdate}
  R_{k,t} = \beta \cdot r_{k} + (1-\beta) \cdot R_{k,t-\gamma}
\end{equation}
For convergence, $0<\beta<1$.
\update{To reduce the weight assigned to the current time interval, for our
experiments in Section~\ref{sec:security}, we use $\beta = 0.2$;
however, we have experimented with other values of $\beta$ ($\le 0.5$) and anecdotally found similar results.}
%
Algorithm~\ref{algo:BMS} summarizes the process of belief maintenance. 
\begin{algorithm}[t]     
\caption{Belief Maintenance System}
\label{algo:BMS}
\begin{algorithmic}[1]
\Procedure{BeliefMaintenance}{$t$}
    \State {\#Risk update using Honey Notification}
    \State \emph{Reinitialize $a_{k}$ \& $c_{k}$ to $0$, for all switch $s_k$}
    \For {\emph{each entry} $e \in$ \emph{Honey Notification} at time $t$} 
        \State $\mathcal{P}_{s,d} \leftarrow getForwardingPath(e)$
        \ForAll {$k \in P_{s, d}$}
            \State \emph{Increment $c_{k}$ }
            \If {\emph{any ATTACK logged in }$e$}
                \State \emph{Increment $a_{k}$ }
            \EndIf
        \EndFor 
    \EndFor
    \ForAll {connected switch $s_k$}
        \State \emph{Update }$r_{k}$ \& $R_{k,t}$           
    \EndFor
\EndProcedure
\end{algorithmic}
\end{algorithm}
 
The BMS creates a ranked list of switches based on their level of suspiciousness (higher $R_{k,t}$ means higher likelihood of being compromised).
This list is a useful resource for the network administrator for remediation or reconfiguration.

\end{techreport}
\section{Security Analysis}
\label{sec:security}

\begin{techreport}

\system creates deception using honey connections from honey hosts representing different enterprise roles.
In this section, we use the PRISM probabilistic model checker to characterize \system's effectiveness against an adversary that is aware of the existence of a \system deployment.
Note that this evaluation assumes \system has raised an alarm.
\update{The evaluation is designed to determine how well \system can identify the compromised switch.}
Recall that our goal is for compromised switches to be ranked as one of the most suspicious.
We begin by presenting our implementation of \system within PRISM and then present the results of the simulation.

\subsection{PRISM Model}
\label{sec:prism}

Probabilistic model checking uses a model construction that represents the behavior of a system over time, i.e., the possible states that the model can be in, the transitions that can occur between states, and information about the likelihood of these transitions~\cite{Kwiatkowska2018}.
It can provide an approximate value of a certain parameter by calculating all possible system paths. 
The PRISM~\cite{PRISM} probabilistic model checker supports three model types: Markov decision processes (MDPs), discrete-time Markov chains (DTMCs), and continuous-time Markov chains (CTMCs).
We chose to use DTMC as it is more realistic for our model to consider time as discrete steps for maintaining the belief state of each switch.

A PRISM model is constructed as the parallel composition of its modules.
The behavior of each module is described by a collection of guarded commands.
\begin{equation*}
[ \hspace{1mm}]\hspace{2mm} guard \rightarrow p_1 : u_1 + \ldots + p_n : u_n ;
\end{equation*}
Here, the guard $guard$ is a predicate over model variables. 
Each update action $u_i$ describes a transition the module can make by giving the variables new values; in the case of DTMCs, $p_i$ is the transition probability. 
\update{If the guard is true, each update is executed according to its probability.
Modules interact with each other through synchronizing on identically labeled commands, thus modules can depend on each other for updates and transitions. }

\begin{table}[t]
  \caption{\system Configuration in PRISM}
  \label{tab:PRISM}
  \begin{tabular}{|c|p{6.0cm}|p{1.1cm}|}
  \hline
  \multirow{5}{*}{\textbf{E}} & \textbf{Environment Features}                & \textbf{Value} \\ \cline{2-3} 
                              & Number of Roles, $E_{\mathit{role}}$                   & 3                   \\ \cline{2-3} 
                              & Number of Rounds, $E_{\mathit{rounds}}$                & 100                 \\ \cline{2-3} 
                              & Number of connections per round, $E_{\mathit{length}}$ & 100                 \\ \cline{2-3} 
                              & Type of Topology, $E_{\mathit{topology}}$              & Fat-Tree~\cite{fatTree}            \\ \hline
  \multirow{5}{*}{\textbf{N}} & \textbf{Nodes}                &  \\ \cline{2-3} 
                              & Network devices or switches, $N_{\mathit{switch}}$                   & 14                   \\ \cline{2-3} 
                              & Number of real client hosts, $N_{\mathit{real}}$                & 50                 \\ \cline{2-3} 
                              & Number of honey client hosts, $N_{\mathit{honey}}$ & 50                \\ \cline{2-3} 
                              & Number of servers, $N_{\mathit{server}}$              & 6          \\ \hline
  \multirow{2}{*}{\textbf{L}} & \textbf{Connectivity}                &  \\ \cline{2-3} 
                              & Forwarding paths, $L_{\mathit{src},\mathit{dst}}$                  &                    \\ \cline{2-3} 
                              & Maximum number of redundancy for each pair, $|L_{\mathit{src},\mathit{dst}}|$          & 8                \\ \hline                             
  \multirow{4}{*}{\textbf{A}} & \textbf{Adversarial Features}                & \\ \cline{2-3} 
                              & Compromised switches, $A_{\mathit{switch}}$            &        $\{1,2\} $           \\ \cline{2-3} 
                              & Target role, $A_{\mathit{role}}$         &              \\ \cline{2-3}  
                              & Attacker confidence on system, $A_{\mathit{confidence}}$ &   \\ \hline  
  \multirow{4}{*}{\textbf{P}} & \textbf{Set of Operational policy}                &  \\ \cline{2-3} 
                              & Connection definition, $P_{\mathit{connection}}$            &                   \\ \cline{2-3} 
                              & Belief maintenance, $P_{\mathit{belief}}$         &              \\ \cline{2-3}  
                              & Attacker actions, $P_{\mathit{attacker}}$ &   \\ \hline  
  \end{tabular}
\end{table}

For modeling complex network behavior using PRISM, we developed a \emph{code generator} that takes in a configuration and outputs PRISM models with necessary modules and transition formulas.
The generated model also 
(1)~ensures consistent state updates and module transitions; 
(2)~identifies compromised switches based on observations from honey connections; 
and (3)~generates the necessary reward functions to measure the performance of the system. 
Our framework generates a dedicated \system model for each configuration. 
Mathematically, each configuration is defined by $\langle E, N, L, A, D, P \rangle$, as described in Table~\ref{tab:PRISM}.
We define three PRISM modules: \emph{Defender}, \emph{System} and \emph{Adversary}. 
The interactions between these modules is summarized in Figure~\ref{fig:PRISM}. 

\subsubsection{Defender Module}
\label{sec:defender_module}

As shown in Figure~\ref{fig:PRISM}, the defender module specifies the current system state by defining a connection $C \rightarrow \langle \mathit{type}, \mathit{role}, \mathit{source}, \mathit{destination}, \mathit{path}\rangle$.
By selecting a new connection configuration, a new transition path is initiated. 
Listing~\ref{lst:code1} shows a simplified segment of PRISM code for selecting a new connection configuration.

Both adversarial actions and the system belief update in the current path depend on the connection configuration. 
Since we cannot represent traffic replication in PRISM, we specify the same probabilistic selection weight for both the honey and real types.
As a result, the model produces a nearly equal number of honey and real connections over the time.

\update{For this implementation, we have only considered three mission-oriented roles, each of which is selected with equal probability.}
The source and destination are randomly chosen for each connection, depending on the type of connection and role chosen in previous states. 
\update{For this PRISM analysis, we have considered both disjoint and non-disjoint paths.
We are using a uniformly random distributed forwarding path selection algorithm.}
Since PRISM cannot directly encode a network topology, our PRISM code generator enumerates these different paths between sources and destinations as distinct PRISM formulas with unique tags.

\begin{figure}[t]
  \centering
  \includegraphics[width=3.3in]{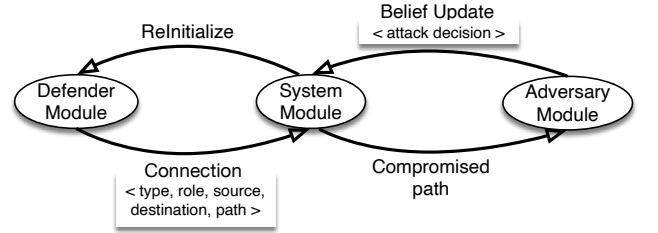}
  \caption{\system workflow between modules in PRISM}
  \label{fig:PRISM}
\end{figure}

\begin{lstlisting}[
  basicstyle=\ttfamily\scriptsize, float=*,caption={$P_{\mathit{connection}}$ code snippet}, label={lst:code1}, frame = single
]
//#type selection# 
[] sched=0 & free & flowType=0 -> (flowRatio/100): (flowType'=1) & (sched'=1)
  + (1-flowRatio/100): (flowType'=2) & (sched'=1);
//#role type selection #
[] sched=1 & free -> 1/3: (roleType'=role0) & (sched'=2) + 1/3: (roleType'=role1) & (sched'=2) 
  + 1/3: (roleType'=role2) & (sched'=2);
//#source selection# 
[] sched=2 & free & sourceID=0 & count<maxIteration & flowType=1 & roleType=role0 -> 
  1/4: (sourceID'=h1) & (sourceSW'=h1sw) & (sched'= 3)+
  ... + 1/4: (sourceID'=h10) & (sourceSW'=h10sw) & (sched'= 3);
...
[] sched=2 & free & sourceID=0 & count<maxIteration & flowType=2 & roleType=role2 -> 
  1/4: (sourceID'=h11) & (sourceSW'=h11sw) & (sched'= 3) +
  ... + 1/4: (sourceID'=h17) & (sourceSW'=h17sw) & (sched'= 3);
//# destination selection#
[] sched=3 & free & destinationID=0 & roleType=role0 -> 1/2: (destinationID'=h21) & (destinationSW'=h21sw)& (sched'= 4) 
  + 1/2: (destinationID'=h24) & (destinationSW'=h24sw)& (sched'= 4);
...
//## current path selection ##
[] sched = 401 & sourceID != 0 & destinationID != 0 -> 1/2 : (currentPath' = pathNumber0) & (sched'=402) 
  + 1/2 : (currentPath' = pathNumber1) & (sched'=402);
\end{lstlisting}

\subsubsection{System Module}
\label{sec:system_module}

The system module gets the current connection $C$ as a configuration. 
It decides between two possibilities.
\update{If the chosen forwarding path contains at least one compromised switch, the system state gives control to the Adversary module. }
Otherwise, 
the system state moves towards the defender module to reinitialize the system.

\myparagraph{Belief Update}
After the adversary module takes actions (Section~\ref{sec:adversary_module}), control returns to the system module.
For every round $r$, the system module records the number of honey connections ($c_k$) handled by each switch $k$, as well as the number of adversarial incidents ($a_k$).
After the completion of each round $c_k$ and $a_k$ are reinitialized.

In our current implementation, each round consists of $E_{\mathit{length}}$ connections (see Table~\ref{tab:PRISM}).
When completing one round, our model goes though approximately $E_{\mathit{length}} \times 20$ (or $\times~25$) state transitions and $E_{\mathit{length}} \times 3$ module transitions.
After completing a round, the current belief is calculated as described in Equation~\ref{eq:crateUpdate}.
Listing~\ref{lst:code2} shows a simplified segment of PRISM code representing this update.
Here, if the current connection type is $honey$ and $attack$ is true, the adversarial incident count ($a_k$) of each switch $k$ on the current $path$ is incremented. 
Note that this code uses $\beta = 0.2$ as discussed in Section~\ref{sec:seceval}.

\begin{lstlisting}[
  basicstyle=\ttfamily\scriptsize, float,caption={$R_{k,t}$ update code snippet from $P_{\mathit{belief}}$}, label={lst:code2}, frame = single
]
// #record honey events#
[beliefUpdateAttacker] sched=5 & active_defender=true &
  (attackFlow=2| attackFlow=3) & (flowType=2) -> 
  (sched'=6) & (received_size'=0); 
...
[] sched=6 & currentPath=3002 -> (ae_sw0'=ae_sw0+1) & 
  (ae_sw4'=ae_sw4+1) & (ae_sw1'=ae_sw1+1) & (sched'=801);
...
[] (flowType=2) & sched=8 & currentPath=3002 -> 
  (count_sw0'=count_sw0+1) & (count_sw4'=count_sw4+1) &
  (count_sw1'=count_sw1+1) & (sched'=(80001));
...
// #update belief# 
[] sched=9 & roundFinished=true -> 
  (bi_sw0'=round((((bi_sw0/100)*0.80) +
  ((ae_sw0/(count_sw0))*0.20))*100)) & 
  ... & (bi_sw7'=round((((bi_sw7/100)*0.80)+
  ((ae_sw7/(count_sw7))*0.20))*100)) & (sched'=10);
[] sched=10 & roundFinished=true -> 
  (count_sw0'=1) & (ae_sw0'=1) & 
  ... 
  (count_sw7'=1) & (ae_sw7'=1) & (sched'=11);
\end{lstlisting}

\subsubsection{Adversary Module}
\label{sec:adversary_module}

\update{To simulate reconnaissance, we assume the adversary receives all possible kinds (different types, roles, IDs) of connections from the defender that pass through the cooresponding compromised switch.}
Note that we assume the adversary has knowledge observed from all compromised switches, if there is more than one.
Ultimately, the goal of \system is to defend against passive reconnaissance by deceiving the adversary with honey flows. In the context of our model, this means the defender cannot detect the adversary or update its belief.
However, an adversary that is aware of \system may still act, performing some active reconnaissance and attacks once it has gained sufficient confidence through passive reconnaissance.
Therefore, the adversary module accumulates confidence in observed information and then attempts to 
(1)~increase confidence through some active reconnaissance, and 
(2)~attack real connections with targeted roles.

As specified in Table~\ref{tab:PRISM}, the adversary module has a target organizational role $A_{\mathit{role}}$ (e.g., IT administrators).
We assume the adversary is only interested in traffic for that role, as defined by connections to the role's corresponding target servers.
The module is also configured with a confidence parameter $A_{\mathit{confidence}}$, which specifies a threshold of sufficient belief in observed information.

The adversary module has two phases:
(1)~attack, and (2)~build confidence. 
For the attack phase, each connection starts with checking whether the current connection as associated with $A_{\mathit{role}}$.
If the current connection matches $A_{\mathit{role}}$, the adversary probabilistically (based on $A_{\mathit{confidence}}$) determines its belief for the current observation. 
If the adversary believes the current observation is real traffic, it performs an attack.
On the other hand, if the adversary believes the current observation is honey traffic, it does nothing. 
Listing~\ref{lst:code3} shows a simplified segment of PRISM code for the adversary module.

\begin{lstlisting}[
  basicstyle=\ttfamily\scriptsize, float=*,caption={$P_{\mathit{attacker}}$ code snippet}, label={lst:code3}, frame = single,
]
confidence: int init 10;
[startAttacker] attacker=0 & active_attacker=true & sched=5 -> (attacker'=1);
[] attacker=1 & active_attacker=true & targetRole=roleType ->
  (confidence/100): (beliefObservation'=true) & (attacker'=2)
  +(1-confidence/100): (beliefObservation'=false) & (attacker'=2);
[] attacker=2 & active_attacker=true & beliefObservation=true & flowType=1-> (attack'=true) & (attacker'=3);
[] attacker=2 & active_attacker=true & beliefObservation=false & flowType=2->(attack'=true) & (attacker'=3);
[] attacker=2 & active_attacker=true & beliefObservation=false & flowType=1->(attack'=false) & (attacker'=3);
[] attacker=2 & active_attacker=true & beliefObservation=true & flowType=2-> (attack'=false) & (attacker'=3);
...
[] attacker=6 & attackerRoundComplete=true & confidence<90 -> 
  2/3: (confidence'=confidence+1) & (attacker'=7)
  + 1/3: (confidence'=confidence-1) & (attacker'=7);
[] attacker=6 & attackerRoundComplete=true & confidence>=90 -> 
  1/2: (attacker'=7)+ 1/2: (confidence'=confidence-1) & (attacker'=7);
[beliefUpdateAttacker] (attacker=7|attacker=6) & attackFlow!=0 & 
  active_defender=true & active_attacker=false -> (attacker'=0);
\end{lstlisting}

On the completion of each round, the adversary probabilistically updates $A_{\mathit{confidence}}$, either increasing or decreasing it. 
To indicate that the adversary's knowledge is increasing with each connection, 
our implementation uses a higher weight (e.g., $\frac{2}{3}$ in Listing~\ref{lst:code3}) for increasing the $A_{\mathit{confidence}}$.
To simulate the effect of deception, we also include the possibility of decreasing confidence (e.g., $\frac{1}{3}$ in Listing~\ref{lst:code3}).
Finally, we assume the adversary cannot have $100\%$ confidence over its observation.
Therefore, if $A_{\mathit{confidence}} \le 90$, our implementation uses an equal probability of increasing or decreasing $A_{\mathit{confidence}}$ (e.g., $\frac{1}{2}$ in Listing~\ref{lst:code3}).
\update{Alternatively, if $A_{\mathit{confidence}} \ge 90$, our implementation only allow it to remain same or decrease.}

\update{Based on this operation, the adversary's action for a connection can be defined by a Markov chain.}
Let \system's initial state be denoted $\langle C_0, A_0, B_0\rangle$, where $C_0$ is the current connection state, $A_0$ is the adversary state in terms of confidence, and $B_0$ indicates system's belief on the suspiciousness of switches.
Figure~\ref{fig:PRISM2} provides a simplified visualization.
The figure assumes only three possible conditions: 
(1)~$compromised$ defines the state of a forwarding path being compromised or not, 
(2)~$type$ defines a connection to be either honey or real, and 
(3)~$attack$ defines an adversarial attack decision.
We assume a connection configuration (e.g., source, destination, type) can repeat; however, this is infrequent and not shown in the figure.

\begin{figure}[t]
  \centering
  \includegraphics[width=3.4in]{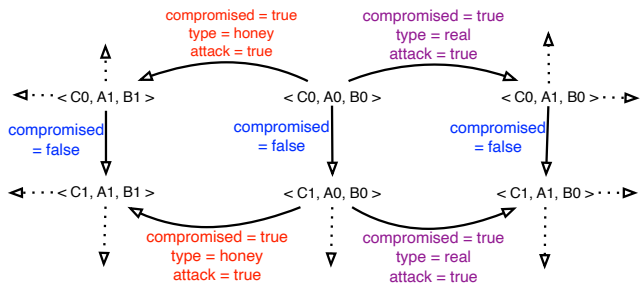}
  \caption{A simplified version of \system Markov chain}
  \label{fig:PRISM2}
\end{figure}

\subsection{Security Evaluation}
\label{sec:seceval}

This section provides the simulation results from the PRISM module described in Section~\ref{sec:prism}.
However, first we describe our experimental setup and performance metrics.

\begin{figure*}[t]
  \centering 
\begin{subfigure}{0.33\textwidth}
\includegraphics[width=\linewidth]{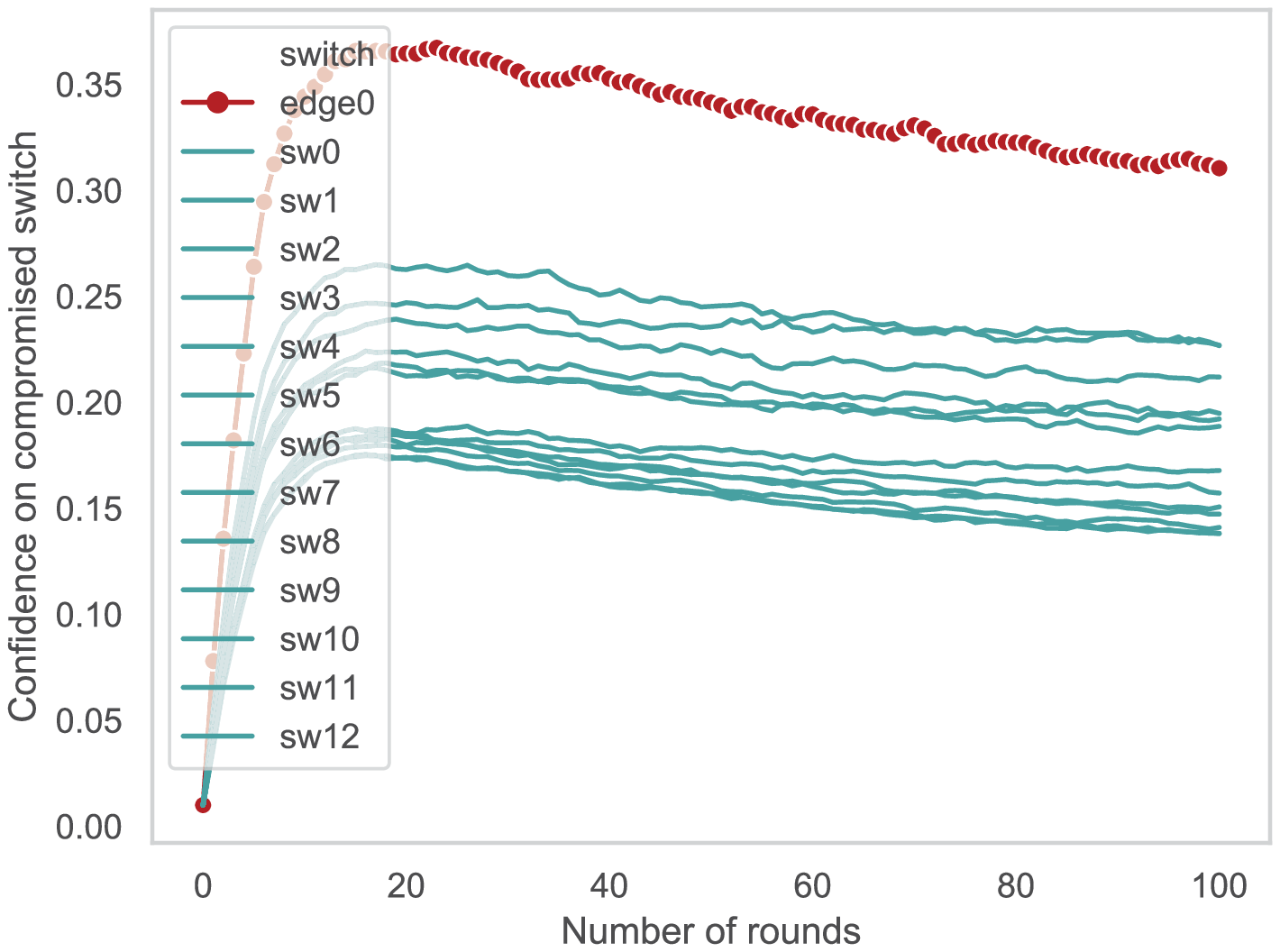}
\caption{One compromised edge switch}
\label{fig:1}
\end{subfigure}\hfill 
\begin{subfigure}{0.33\textwidth}
\includegraphics[width=\linewidth]{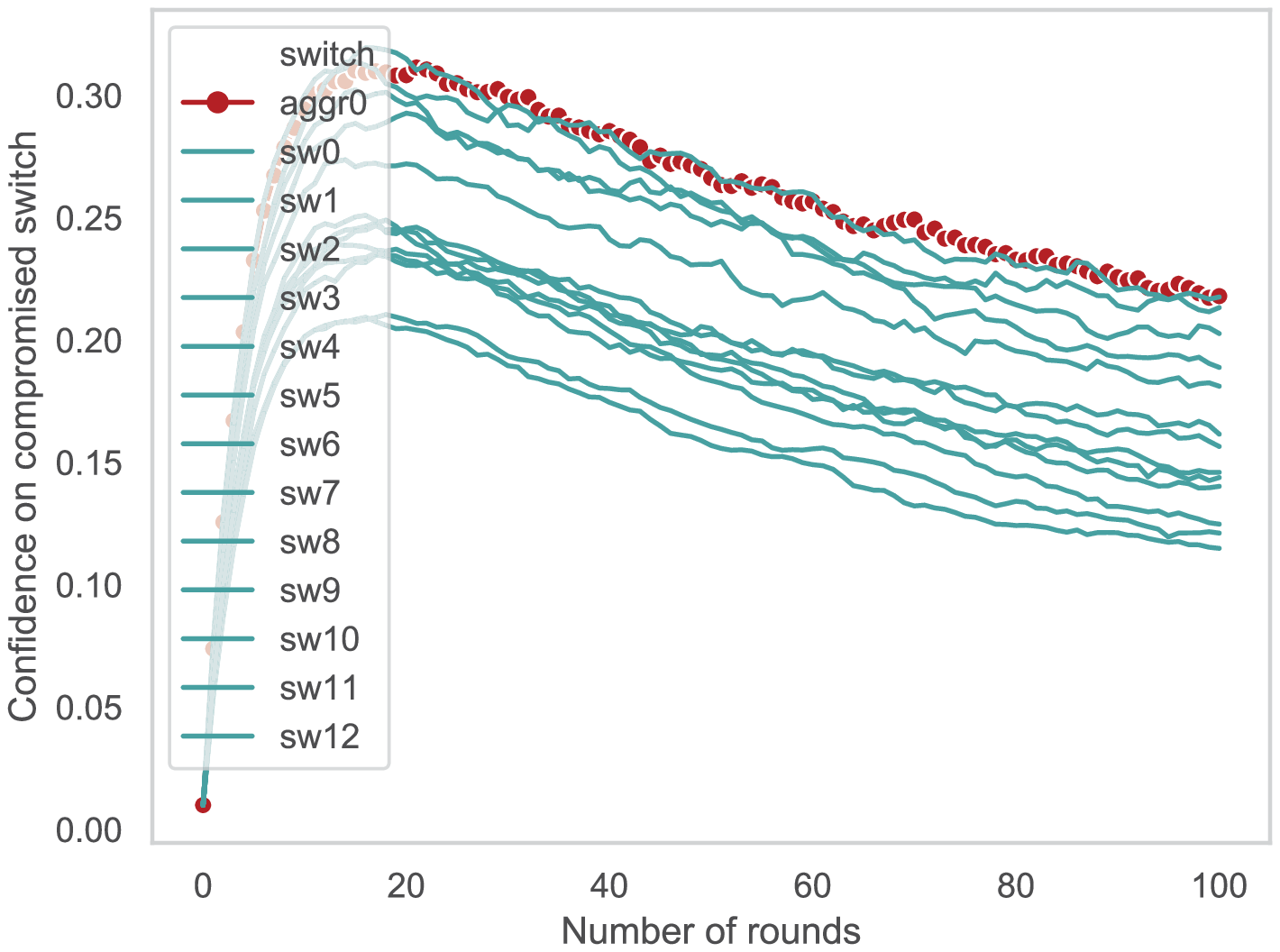}
\caption{One compromised aggregate switch}
\label{fig:2}
\end{subfigure}\hfill 
\begin{subfigure}{0.33\textwidth}
\includegraphics[width=\linewidth]{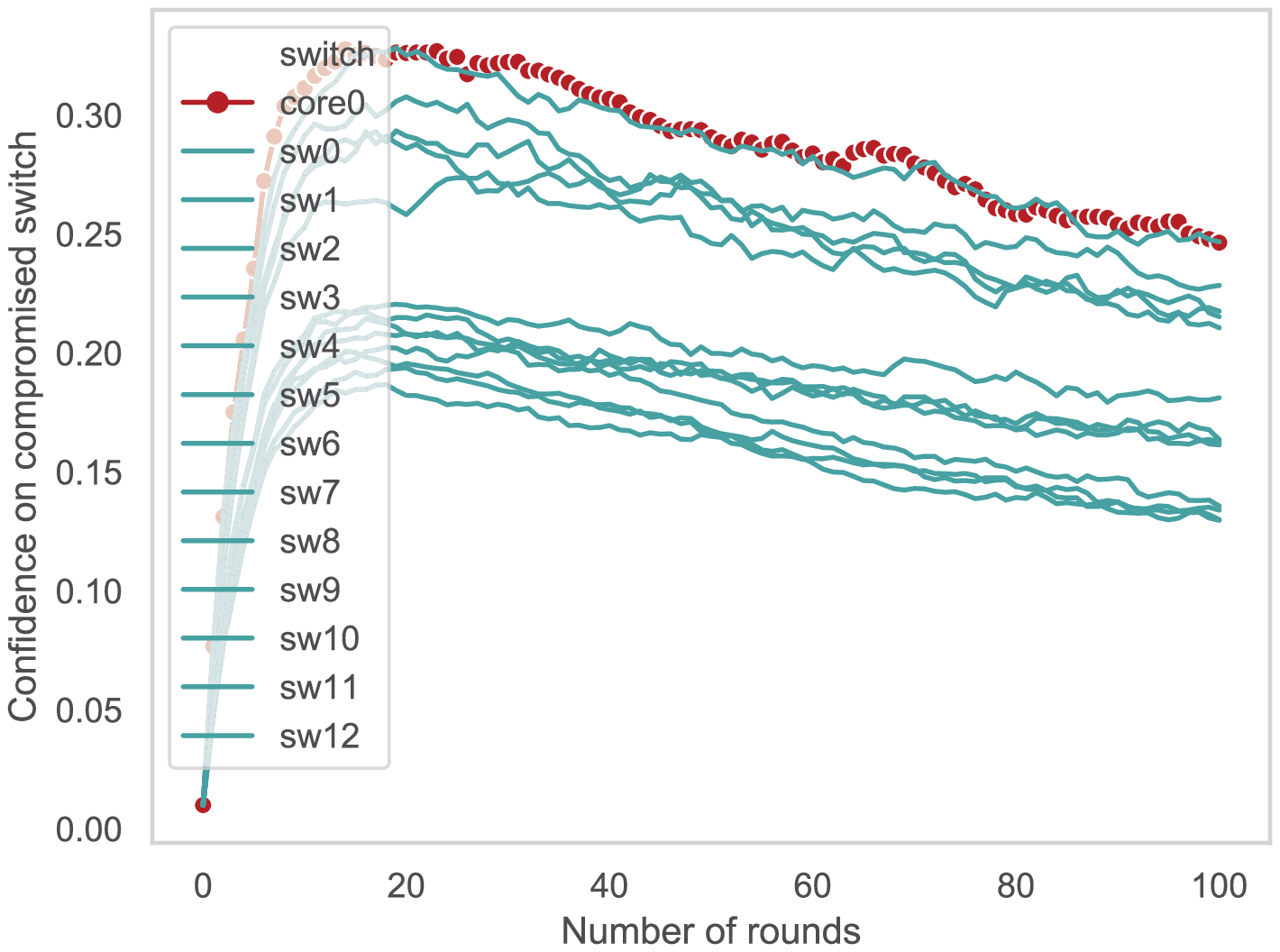}
\caption{One compromised core switch}
\label{fig:3}
\end{subfigure}
\caption{Confidence in switch compromise for one compromised switch ($\beta = 0.2$).}
\label{fig:confidence1}
\end{figure*}
\begin{figure*}[t]
\centering 
\begin{subfigure}{0.33\textwidth}
\includegraphics[width=\linewidth]{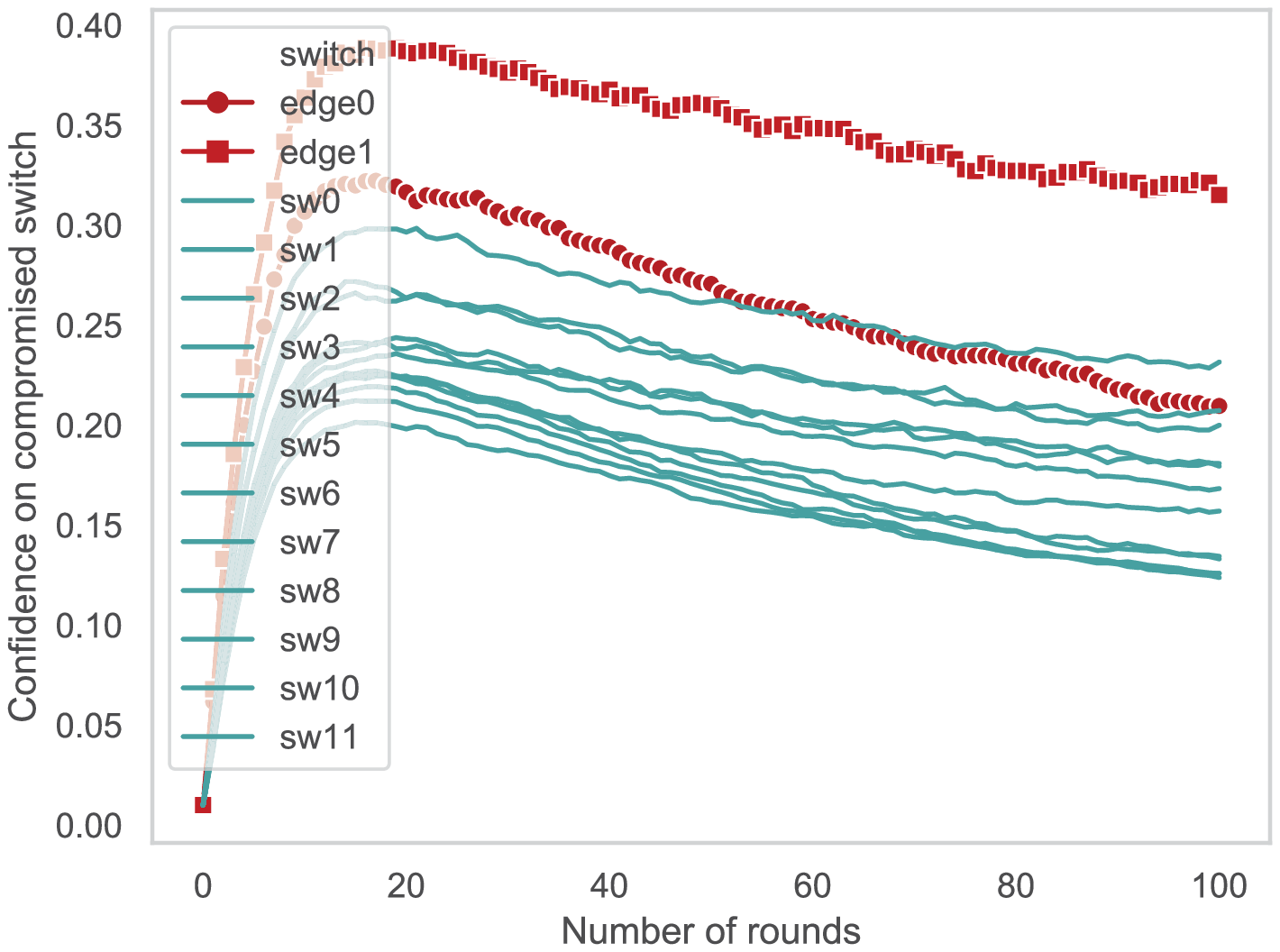}
\caption{Two compromised edge switches}
\label{fig:4}
\end{subfigure}\hfill 
\begin{subfigure}{0.33\textwidth}
\includegraphics[width=\linewidth]{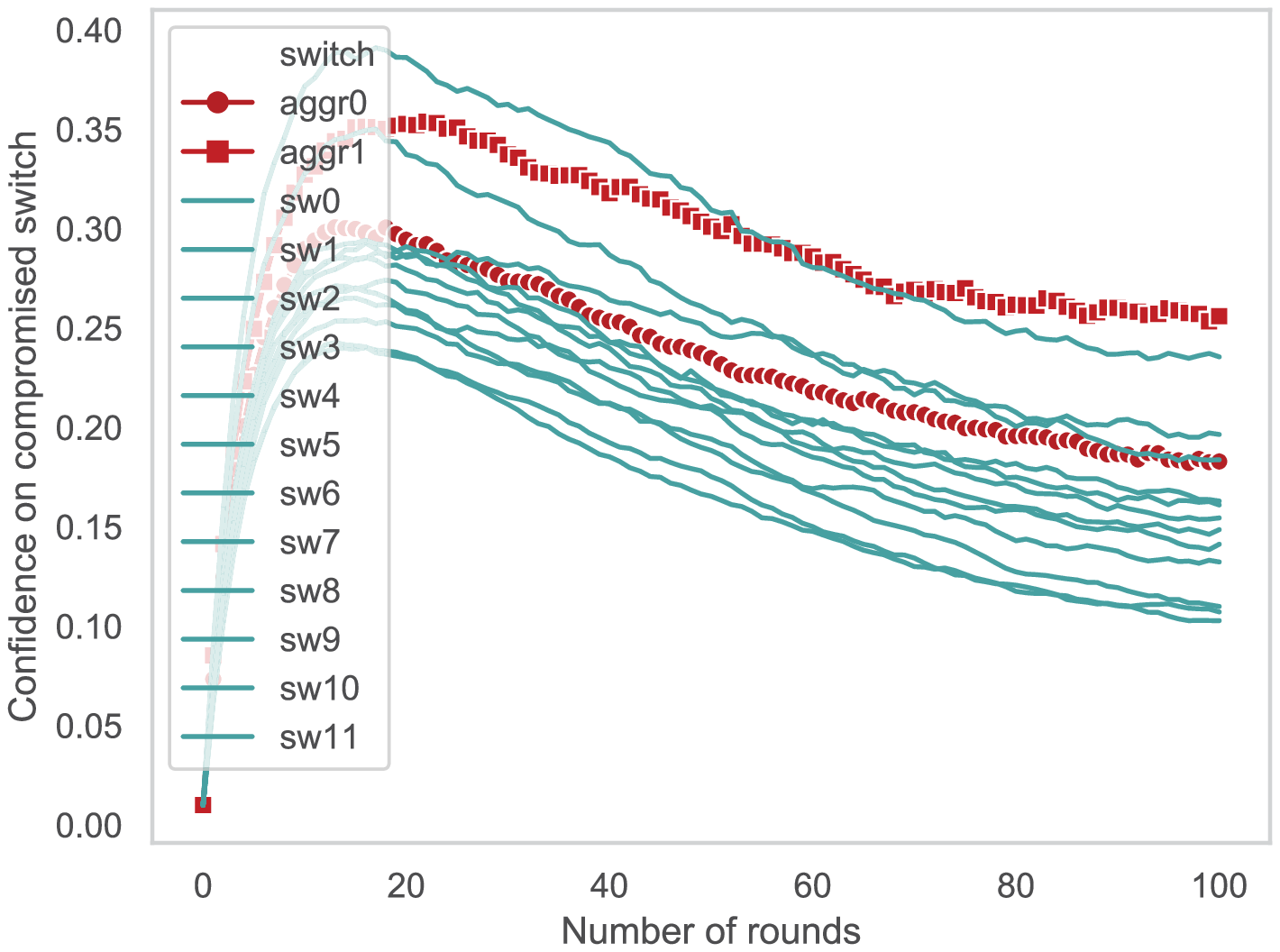}
\caption{Two compromised aggregate switches}
\label{fig:5}
\end{subfigure}\hfill 
\begin{subfigure}{0.33\textwidth}
\includegraphics[width=\linewidth]{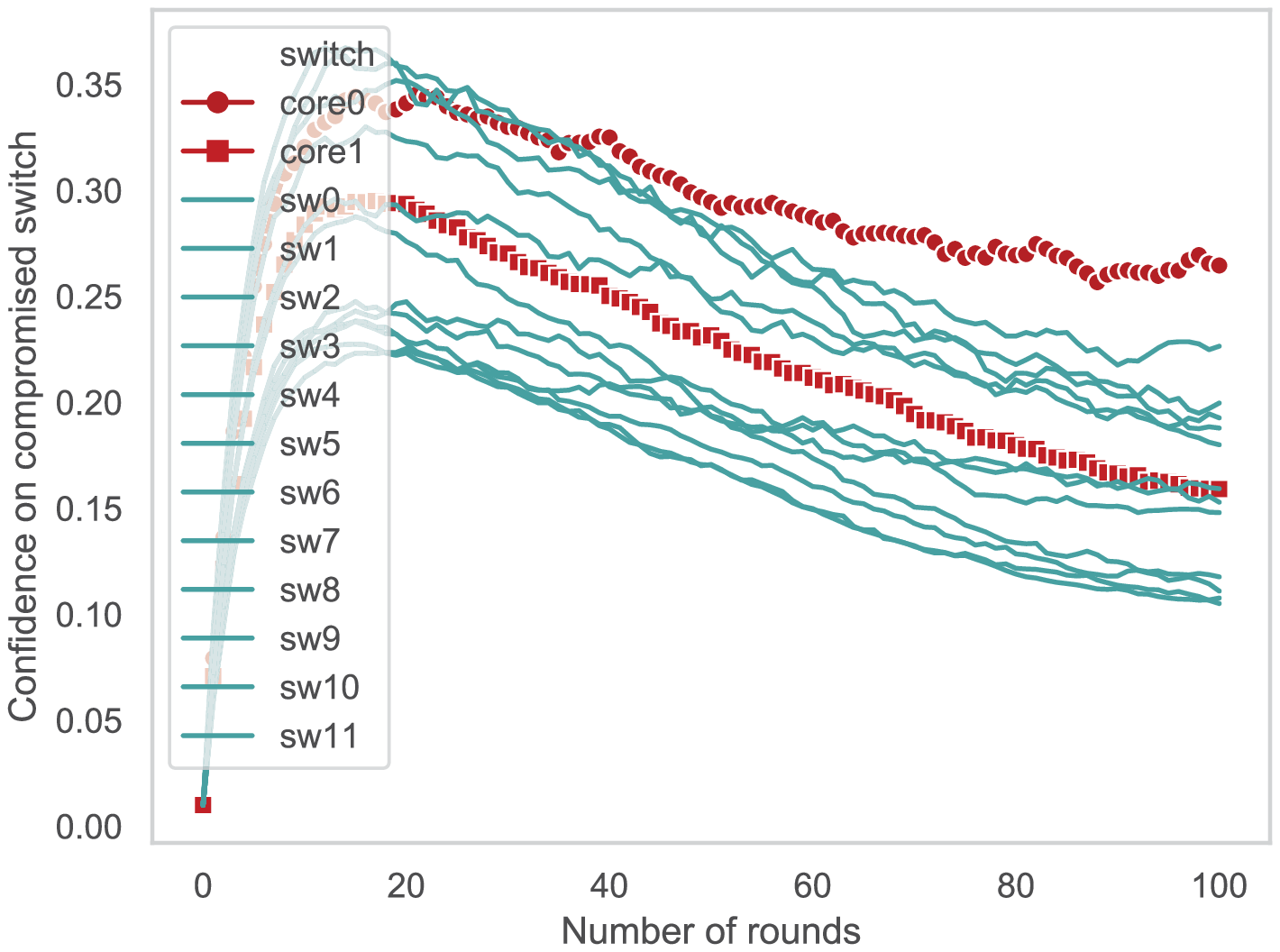}
\caption{Two compromised core switches}
\label{fig:6}
\end{subfigure}
\begin{subfigure}{0.33\textwidth}
\includegraphics[width=\linewidth]{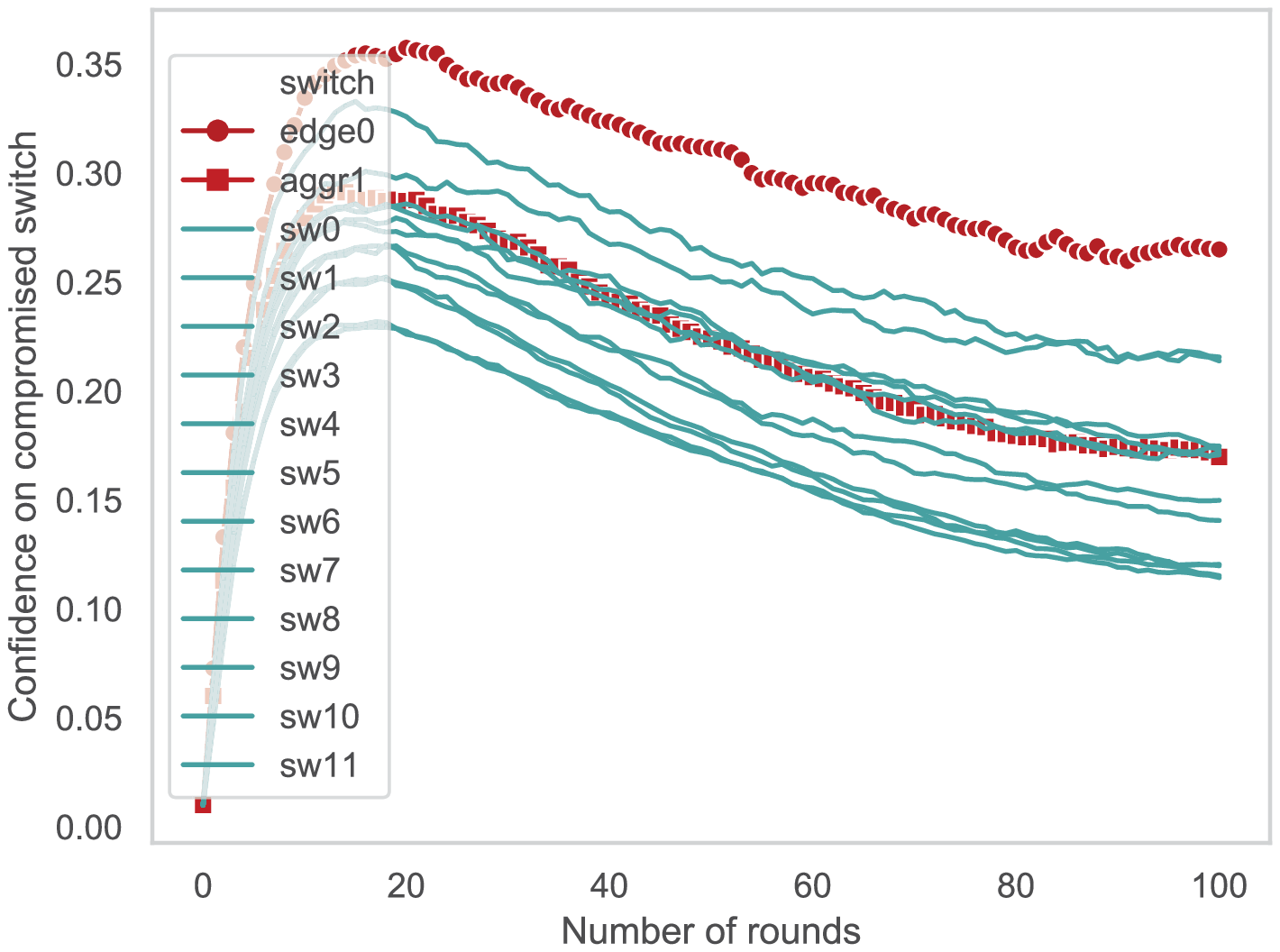}
\caption{One edge and one aggregate compromised switch}
\label{fig:7}
\end{subfigure}\hfill 
\begin{subfigure}{0.33\textwidth}
\includegraphics[width=\linewidth]{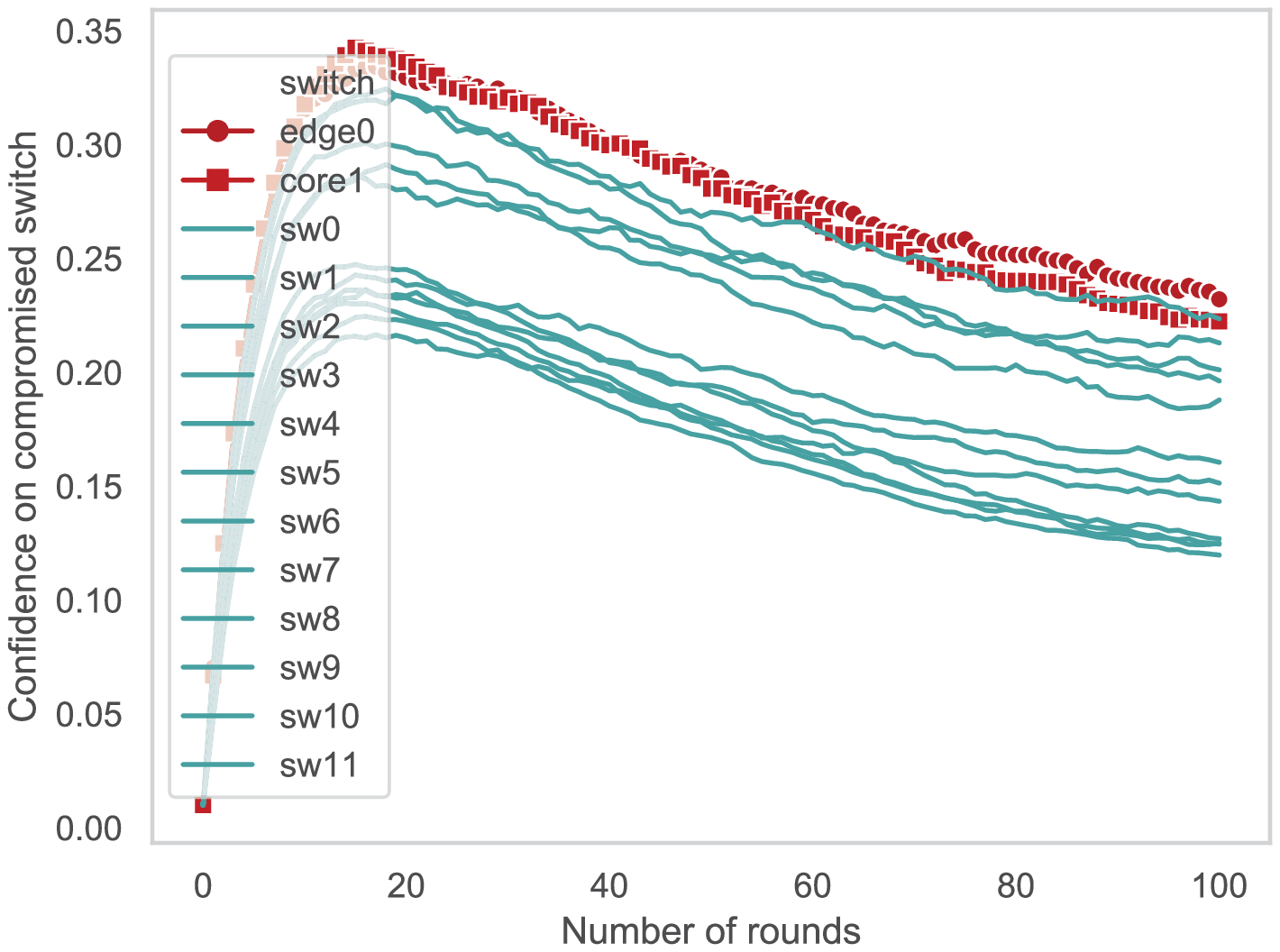}
\caption{One edge and one core compromised switch}
\label{fig:8}
\end{subfigure}\hfill 
\begin{subfigure}{0.33\textwidth}
\includegraphics[width=\linewidth]{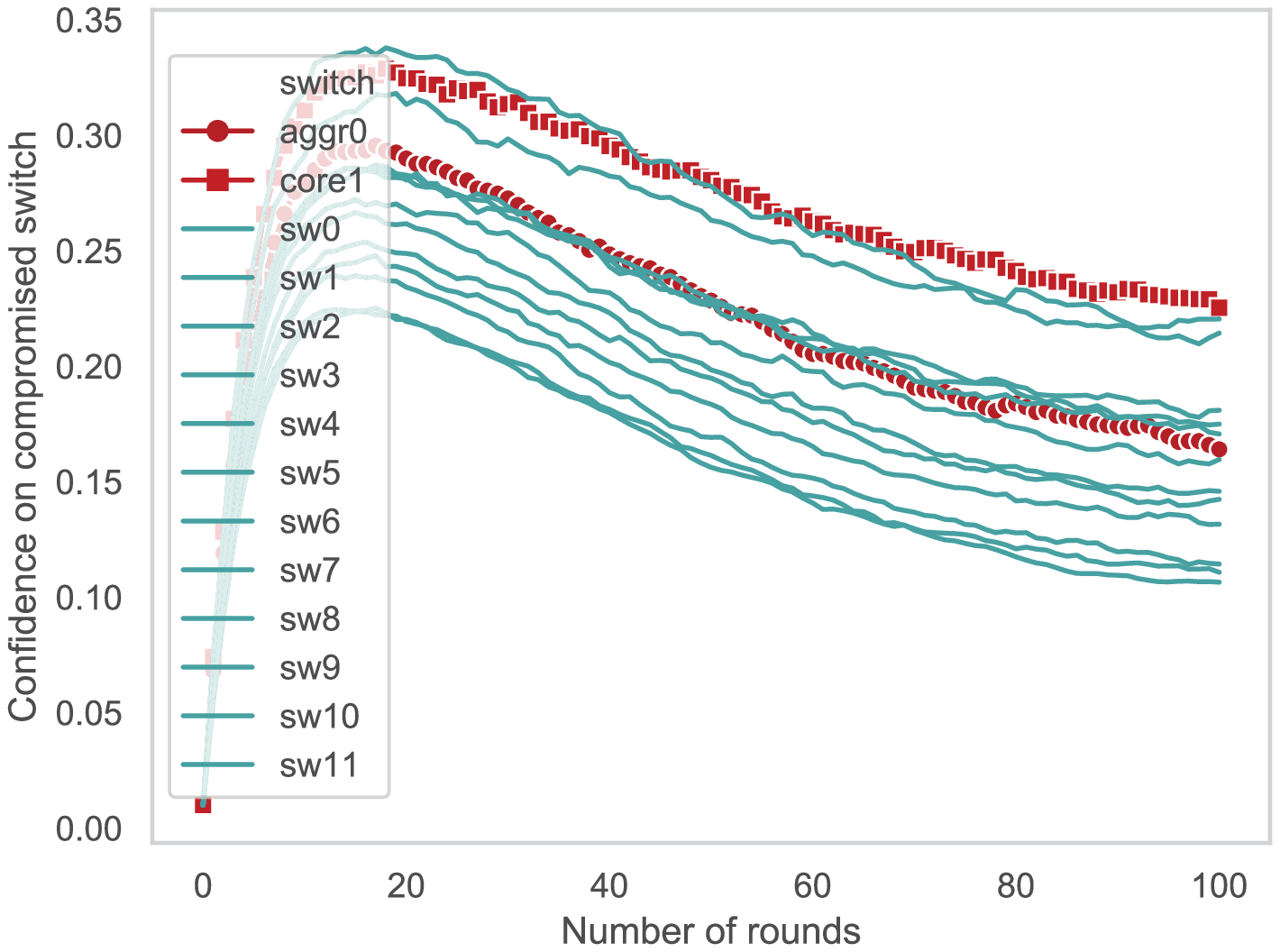}
\caption{One aggregate and one core compromised switch}
\label{fig:9}
\end{subfigure}
\caption{Confidence in switch compromise for two compromised switches ($\beta = 0.2$).}
\label{fig:confidence2}
\end{figure*}

\myparagraph{Experimental Setup}
As described in Section~\ref{sec:prism}, our code generator automatically creates a PRISM model given a system configuration.
The code generator was written in around 1,300 lines of Python code.
It has two parts: 
1)~the \emph{TopologyParser} generates the topology by using connectivity information to enumerate all-possible forwarding paths for each pair of edge switches, 
2)~the \emph{PRISMCodeGenerator} takes the topology information and the system parameters (Table~\ref{tab:PRISM}) and generates final PRISM logic.
The ``Experiment'' column in Table~\ref{tab:PRISM} specifies the configuration used for our experiment.
Specifically, we considered scenarios where there were 1 or 2 compromised switches, including simulations where the compromised switch resided at different locations within the Fat-Tree topology (i.e., edge, aggregate, core).
Note that we used a Fat-Tree topology for lack of a public database of an enterprise network topology.
Repositories such as Topology-Zoo~\cite{TopoZoo} and Internet2~\cite{internet2} only include topologies for data centers, ISPs, and point of presence (POP) networks.
However, our code generator can consume topologies in Geography Markup Language (GML) following the format of Topology-Zoo and can therefore be easily used to evaluate different topologies.
Finally, to assess the sensitivity of $\beta$ in Equation~\ref{eq:crateUpdate}, we ran the simulator with $\beta \in \{0.1, 0.2, 0.3, 0.4, 0.5\}$.


We used the discrete-event simulator built into PRISM, a technique often called statistical model checking~\cite{SMC}. 
This sampling approach generates a large number of random paths through the model, evaluating the result of the given properties on each run, and using this information to generate an approximately correct result~\cite{PRISM}. 
Each simulation was run for the path length of $200,000$ (approximately $100$ rounds) and collected data on each step of $2,000$ (approximately on completion of each round). 
Each simulation takes $50$ samples and provides the mean values as a final result.

\myparagraph{Performance Metrics}
Recall that this evaluation is designed to determine how well \system can locate the compromised switch or switches, i.e., how often the compromised
switch(es) appear high in \system' suspiciousness ranking.
We thus examine this ranking over the course of $100$ rounds.


\subsubsection{Detection Accuracy with One Compromised Switch}\label{subsubsec:oneComp}

Figure~\ref{fig:confidence1} shows the relative ranking of suspiciousness for switches for $\beta = 0.2$ when there is only one compromised switch.
The other $\beta$ configurations produced anecdotally similar graphs, but as hypothesized, a smaller $\beta$ performs better.
The figure shows that when there is one compromised switch, that switch is consistently ranked in the top-1 or top-2.

When comparing the different locations for the compromised switch (i.e., edge, aggregate, and core),
the figure shows the best performance for compromised switches located at the edge (Figure~\ref{fig:1}).
This is because it is easier to isolate the attack activity over the time.
As shown in Figures~\ref{fig:2} and~\ref{fig:3}, when a core or aggregate switch is compromised, \system does not provide as clear of a distinction.
However, this is an artifact of the Fat-Tree topology, as core and aggregate switches are included in most of the forwarding paths that raise alarms.
Hence, it is difficult to statistically determine which switch on the path is performing the attacks.
That said, even with this high overlap, the compromised switches were within the top-2 riskiest at all times.

\subsubsection{Detection Accuracy with Two Compromised Switches}

Figure~\ref{fig:confidence2} shows six possible combinations of compromised switches for $\beta = 0.2$.
Other than the number of compromised switches, the other parameters remained the same as in the tests with one compromised switch.
As before, different $\beta$ values produced visually similar results, with smaller $\beta$ values performing better.

\update{As discussed in Section~\ref{subsubsec:oneComp}, edge switches
are easier to isolate than aggregate and core.
When aggregate or core switches are compromised, at least one of the compromised
switches is ranked in the top one or two most of the time, with the
second compromised switch being in the top five for all but two scenarios.}
Note that the network administrator can approach refreshing switches to a good known state in an incremental fashion.
That is, it can refresh the top-1 switch, removing one of the compromised switches and leaving only one, which as shown in Figure~\ref{fig:confidence1} is easier to isolate.
While the adversary will know that it has been detected, in the worst case (for detection) it will stop attacking connections, which is ultimately our goal.
\update{The system administrators can also define some threshold on the switch risk factors, depending on their security requirement.
Thus, administrators will remove a switch only when its risk factor goes beyond that threshold.}
  
\end{techreport}
\section{Performance Evaluation}
\label{sec:eval}

\begin{techreport}
\system's security stems from its deception elements (e.g., honey connections and routes), which add network overhead.
We now discuss our prototype implementation, experimental setup, and evaluate \system's performance overhead in an emulated Mininet~\cite{Mininet} environment.

\subsection{Implementation} \label{subsec:imple}
Our \system prototype is implemented as six components that comprise the design in Section~\ref{sec:design}. 
We built our prototype on top of the OpenJDK 11.0.7 and ONOS 2.0.0 SDN controller with the default configuration. 
Three components are implemented as ONOS Java applications: ForwardingPath Manager (95 lines of code), Heartbeat Generator (305 lines of
code), and Traffic Tracer (250 lines of code). 
Two additional components run as dedicated processes that communicate with the ONOS controller: Belief Management System (180 lines of code) and Honey Connection Processor (310 lines of code).   
The Mininet network creation and real host traffic generator took up 600 and 120 lines of code, respectively.
The Honey Agent (240 lines of code) is used implementation the workflow of a honey host.
\update{Although \system can function with different applications (e.g., SSH, SMTP), we restricted ourselves to HTTP/HTTPS traffic}.

\begin{figure}[h]
  \centering 
  \includegraphics[width=0.90\columnwidth, clip, trim=0.05in 0.05in 0.05in 0.02in]{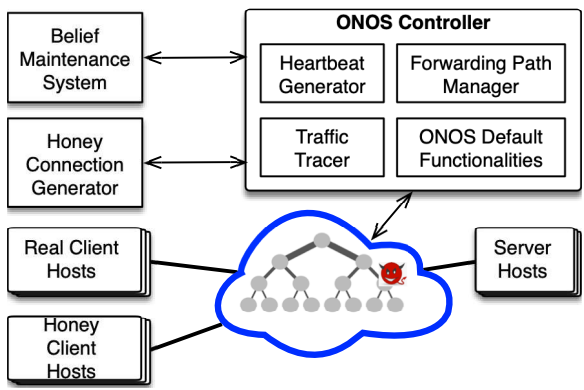}
  \caption{Experimental Layout for Evaluation}
  \label{fig:layout}
\end{figure}

\myparagraph{Network Creation}
Our performance analysis uses the same Fat-Tree topology generation algorithm used for the security analysis in Section~\ref{sec:security}. 
Both \emph{Real} and \emph{Honey} client hosts are implemented as standard Mininet hosts. 
The Real hosts execute a script that randomly initiates sessions (a sequence of one or more HTTP requests) with a target server. 
The Honey hosts execute the Honey Agent script, which uses \emph{heartbeat} instructions from the controller to initiate a session with a target server and then reports results back to the controller using \emph{Honey Notifications}. 
Servers are implemented as Docker containers running on the host machine. 
To enable SSH and ping traffic between the servers and Mininet, we created a virtual node in the root network namespace and then created a link between this node and one of the core switches in the Mininet topology.

\subsection{Experimental Setup}
\update{The evaluation was hosted in a virtual machine configured with 8 vCPUs and 32 GB RAM, running on a VMware ESXi 6.5.0 host with Intel(R) Xeon(R) CPU E5620 @ 2.40GHz processors.
Figure~\ref{fig:layout} shows the network, along with the main components described in the implementation details (Section~\ref{subsec:imple}).}
We considered four environments to compare the performance impact of various \system components.
The \emph{Baseline} environment was configured to use the default ONOS settings with no \system features enabled and included only real hosts in the network. 
The \emph{Honey Forwarding} environment replaces the default ONOS reactive forwarding application with the \system Forwarding application but does not introduce honey hosts or honey agents. 
The \emph{Honey Host} environment was configured with the default ONOS forwarding app but introduces the Heartbeat Generation Application and honey hosts. 
Here, we have one honey host initiated in correspond to each real host in the network.
Finally, the \system environment was configured with all \system features enabled and one honey host for each real host in the network.

\update{To maintain consistency with our security analysis, we configured each environment with 50 real hosts.
As discussed easier in Section~\ref{sec:design}, we are using a 5-tuple for flow rule matching: $s_{ip}$, $s_{port}$, $d_{ip}$, $d_{port}$, and $protocol$.
The baseline and all treatments use ONOS's default 10 second idle timeout for flow-mod rules.
Each experiment ran for 30 minutes and all hosts (honey or real) were configured to send 1 request per second to a specific server, which is selected based on their roles from a fixed set. 
This network load represented a moderately loaded network, as introducing honey connections at peak capacity would clearly have a significant overhead.
We assume enterprise networks rarely operate at a peak capacity for prolonged durations.
Throughout the experiment, the request completion time was recorded for every request between unique real-client and server pairs.}

\begin{figure}[t]
  \centering 
  \includegraphics[width=0.9\columnwidth,clip,trim=0.1in 0.10in 0.625in 0.35in]{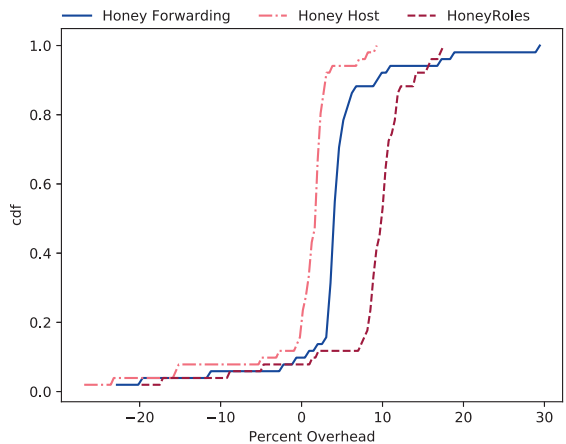}
  \caption{Percent Overhead of HTTP request completion time in each configuration compared to baseline.}
  \label{fig:peroverhead}
\end{figure}

\subsection{\system Performance Overhead}

\update{
  Calculating a single average overhead across all pairs does not provide a useful characterization, as different pairs have different numbers of hops between them, resulting in a significant variance in completion time.
Therefore, to observe the overhead \system imposes on real network traffic,
 we calculated the average request completion time between each unique real-client server pair for the baseline environment.
For each non-baseline environment, we calculated the percent overhead of every real request from the baseline average. 
We plotted the percent overheads for each configuration as a cumulative distribution function (CDF).}

\update{
Figure~\ref{fig:peroverhead} depicts the overhead of each treatment with respect to the baseline.}
Each line represents the percentage of real requests in each environment that finished under a given percent overhead calculated using the average baseline request completion time for unique client-server pairs.
That is, each request between client $c$ and server $s$ in the Honey Forwarding Application, Honey Host, and \system environments was compared to the average completion time of all requests between $c$ and $s$ in the baseline environment.
From this graph we observe that for \system, \update{90\% of requests finish with less than 14\% overhead when compared to the baseline.}

\update{Note that these percentage overheads are for small request-completion times, which are significantly impacted by jitter.}
The median request completion time in the baseline environment was 31 ms.
As a result, even small changes in completion time in the other environments show as a larger magnitude overhead.
For example, with a 31 ms baseline completion time, a request with a 9 ms increase from the baseline (e.g., $40$ ms) results in a $29\%$ overhead.
The natural jitter in network requests and the sensitivity when dealing with small numbers can also explain the negative overheads observed in Figure~\ref{fig:peroverhead}.

\update{Further, we compared the average request-completion time of each environment to the baseline average by calculating the effect size using \emph{Cohen's d}.
Cohen's d reports how many pooled standard deviations two groups differ by.
According to Cohen~\cite{cohen:1988}, a $d$ value of 0.20 is considered a ``small'' effect, a $d$ value of 0.50 is a ``medium'' effect, and a $d$ value of 0.80 is a ``large'' effect of an experimental change to a control group.
For the \system environment, we observe 55\% of client-server pairs have a $d$ value of under 0.20 (small effect), and all of $d$ values are below 0.63, which is well below the 0.80 margin (large effect).
Although, Cohen does not describe how to interpret $d$ values just above a threshold, we found that 90\% of client-server pairs have a $d$ value under 0.26, which we consider to still be relatively small.
Thus we observe that, with respect to request-completion times, \system had a small effect for a majority of the client-server pairs and a medium effect for all the rest.}

We believe our small overheads are due to two primary reasons.
First, the network is not under full load, thus the introduction of Honey Host traffic does not compete with real traffic for resources and has minimal impact on the network links and server processing.
Second, although the Honey Forwarding application may select non-optimal routes for traffic, for the choice of Fat-Tree the non-optimal routes do not introduce major differences in request completion time. 
It may be possible that a network is designed in such a way that a non-optimal route may introduce much higher round trip times but these routes are not permanent and some traffic will still travel over optimal or close to optimal routes.
If this is observed it practice, network administrators can create additional network links to provide shorter alternative paths.


\end{techreport}
\section{Discussion}
\label{sec:discussion}

\begin{techreport}

\todo{\begin{itemize}
    \item Can we disscuss about different types of attack scenarios that can be addressed by HoneyRoles?
    \item Should we discuss the benefits we can get through the defensive deception? We can list some malicious actions which are difficult to mitigate or detected via the traditional defensive technology.
\end{itemize}}

\myparagraph{Attack variations}
\system considers that an adversary uses \emph{passive} and
\emph{active} reconnaissance to obtain knowledge about target
enterprise roles, presumably to launch active attacks using that knowledge.
Many active attacks and reconnaissance techniques have been discovered over the past decades.
\update{Although we illustrate \system with SSL-stripping and blackholing attacks, it is possible to adapt to other types of attack though implementing respective detection algorithms in \emph{honey agents}.}
For example, consider the use of probing messages during active reconnaissance to get a response from target hosts. 
\update{In this case, it is possible incorporate unexpected connections to the IP addresses of honey hosts.}

\myparagraph{Accuracy vs. deception} 
Using the centralized control of SDN, it is possible to dynamically change honey components according to system belief to improve accuracy.
The \system forwarding path manager could route most real connections through less risky switches and honey connections through more risky switches in order to better identify the compromised switch.
However, such an approach would be risky.
First, sudden changes in system behavior may alarm the adversary and
reduce the effectiveness of the deception (e.g., helping it identify
which IP address belong to honey hosts). 
Second, a dynamic change in
system behavior may increase complexity in large networks.  
\update{Third, the TCB must include
at least a segment of switches to achieve the security goal.}

\myparagraph{Scope of implementation}
We used PRISM to evaluate the security of \system and used an emulated Mininet environment to measure performance overhead.
These evaluation frameworks are approximations of realistic enterprise networks.
Our security evaluation was limited in the way it modeled attacker behavior, as we could not find any realistic attack data for enterprise reconnaissance.
Absent realistic attack behavior, the PRISM model was more
comprehensive than a Mininet simulation to estimate detection accuracy.
Additionally, we were unable to find realistic enterprise network topologies and relied on the Fat-Tree topology as a representative topology with redundant links and switches.

\update{Finally, as stated in Section~\ref{sec:overview}, we assume the existence of an ambient network traffic generator~\cite{Swing:Vishwanath:2009, Sommers:2004:SNT, Bowen:2012:SGI}, which our implementation does not include.
Additional work is required to design and evaluate ambient network traffic generators against more recent machine learning algorithms; however, doing so is orthogonal to the contributions of this paper.
Any viable traffic generator could be easily incorporated into \system.
We also note that some machine learning algorithms require significant storage and computational capabilities, which are not available to an adversary positioned on a compromised packet forwarding device.}


\end{techreport}

\section{Related Work}
\label{sec:relwork}

\begin{techreport}
\myparagraph{Network reconnaissance and eavesdropping}
\update{
    Traditional intrusion detection systems cannot detect passive attackers performing network reconnaissance from compromised packet forwarding devices.
    Such reconnaissance investments are particularly apropos to advanced persistent threats (APTs)~\cite{chen2014study}.
    After gaining a foothold in a network by compromising a device, adversaries can leverage information through reconnaissance to better identify targets and vulnerable devices.
    Bartlett et al.~\cite{bartlett2007understanding} demonstrate the dangers of reconnaissance by presenting a quantitative comparison and evaluation of the effectiveness of passive monitoring and active probing for service discovery in decentralized networks.
}

\update{   
    Even if traffic is encrypted, reconnaissance remains a threat.
    Schuster et al.~\cite{Beauty}, Backes et al.~\cite{side}, and Ling et al.~\cite{delay} show that encrypted web traffic can leak information through packet length, packet timing, web flow size, and response delay.
    Similarly, AppPrint~\cite{miskovic2015appprint} analyzes the possibility of fingerprinting mobile apps via comprehensive traffic observations. 
    Anderson et al.~\cite{TSL:2019} have produced TLS fingerprint from network data, revealing the details of TLS versions and other configuration parameters.
    With increasing threats of targeted reconnaissance and attacks (e.g., Snowden~\cite{snowden:2014}, CISCO SYNfulKnock~\cite{SYNfulKnock:2015}, political espionage~\cite{Li:2011:APT}), defense against APT is becoming more critical. 
}

\myparagraph{Deceptive Defenses}
Deception techniques provide alternative defense approach that can mislead and delay adversarial efforts, and even detect attacks in early stages.
Whaley et al.~\cite{whaley1982} define deception as the misperception that is intentionally induced by other entities. 
Spafford et al.~\cite{almeshekah2016} extend this definition of cyber-deception as ``planned actions taken to mislead and/or confuse attackers and to thereby cause them to take (or not take) specific actions that aid computer-security defenses.'' 
Current deceptive defense techniques primarily use a moving target defense (MTD) approach.
These solutions depend on mimicking random or static specification of system behavior, network configuration, or network infrastructures (e.g., honeypots, honey-nets)~\cite{ParrotDead,deception:2006, Jafarian2015, Honeypots}.
Dynamically generated decoy traffic~\cite{decoy2012}, decoy based IP randomization~\cite{decoyIP}, and decoy vulnerability-based honey traffic~\cite{anjum2020optimizing} has reported to be more effective.

\update{
The dynamic control and programmability of an SDN environment has inspired new deception techniques.
HoneyMix~\cite{han2016honeymix} uses a dynamic SDN-based honey-net to automate interactions with adversaries, and showed deception is a promising approach toward defending against network reconnaissance.
Further, the dynamic network configuration of an SDN can be used for discriminating against scanning attacks and enhancing targeted defenses~\cite{adaptingHop,anjum2020optimizing}.
For example, Achleitner et al.~\cite{Achleitner:CyberDeception:2016} use SDN to defend against insider reconnaissance by simulating virtual network topologies as decoys.
}

\todo{We have to update the related work section keeping the R2 comments in mind. Mostly we have to include an comparative study between those literature and HoneyRole}

\myparagraph{Software Defined Networking} \update{SDNs have the potential to address many operational and security challenges in large enterprise networks~\cite{Levin:2014:Panopticon, Lorenz:2017:SDN-Enterprise}. 
They decouple network control from the underlying data plane and consolidates configuration to a logically central controller, which provides valuable flexibility for programmatically and dynamically reconfiguring traffic forwarding~\cite{McKeown:2008:OpenFlow}.
SDN has the potential to supplant conventional security systems~\cite{YOON:2015:SDN}, simplify policy enforcement~\cite{Qazi:2013:SMP,Goutam:2019:HSL}, ensure information flow control~\cite{OConnor:2018:PivotWall}, enable deceptive defense~\cite{HoneyScope:2019}, and provide a software defined perimeter~\cite{SDP,Nayak:2009:RDA}. 
However, the greater capabilities and open functionality of SDN switches increase the potential for compromise and enable a new vantage point for attacks, e.g., data plane attacks using advanced reconnaissance, data manipulation, and redirection (e.g., Teleportation~\cite{Thimmaraju:Cloud:2016}, Benton et al.~\cite{Benton:2013:OVA}, Menghao et al.~\cite{Zhang:reflection:2018}, Know Your Enemy~\cite{Conti_2017}).
} 

\update{
Network analysis and auditing tools (e.g., Header Space Analysis~\cite{HSA:2012}, VeriFlow~\cite{VeriFlow:2013}, SDN-RDCD~\cite{SDN-RDC:2018}) can protect against network or SDN controller configuration failures (or attacks). 
However, a compromised SDN data plane can introduce different types of attack scenarios~\cite{Markku:2014:Spook}, which are not possible to detect through header flow analysis alone. 
Some solutions have sought to detect forwarding attacks by monitoring flow statistics from neighboring switches~\cite{FlowMon}, verifying OpenFlow events in the controller~\cite{Zhang:reflection:2018, Wang:2015:FloodGuard}, applying heavy-weight cryptographic approaches~\cite{Lightweight:Kim:2014}, and naive controller generated probes~\cite{detectCompromisedSwitch:2015, Chao:2016:SDNData}.
}

\update{
Sphinx~\cite{dhawan:2015:sphinx} uses SDN control messages for incremental validation of network updates and detect suspicious behaviors (e.g., DoS, blackholing, fake topology).
WedgeTail~\cite{shaghaghi:Wedgetail:2017} detects both forwarding attacks and forged packets by utilizing Header Space Analysis and other network troubleshooting tools.
Both Sphinx and WedgeTail dynamically construct network flow graphs to compare with a defined policy to identify deviations, which is not only a manual and error-prone process but also cannot handle dynamic networks.
Additionally, DynaPFV~\cite{Li:2018:Dyna} proposed a mechanism to detect packet-modification by comparing the cryptographic hash of packets at the ingress and egress points of a network.
However, none of these prior works can address passive (or even subtle active) reconnaissance. 
Since reconnaissance can be performed without network disruption attacks (e.g., forwarding, packet forging, and packet-modification attacks detected by the tools above), the attacker is able to evade the defenses of prior works. \system complements the detection capabilities of prior works by adding a layer of deception to lower the effectiveness of reconnaissance in the network.
}  
\end{techreport}

\section{Conclusion}
\label{sec:conc}

\begin{techreport}
The increasing complexity of packet forwarding devices such as routers and switches make them a new target for advanced persistent threats.
From the vantage point of a compromised packet forwarding device, an adversary can passively monitor network traffic to identify not only the network topology and servers listening on ports, but also the client hosts that connect to high-value servers such as domain controllers and financial systems.
In this paper, we presented \system as a novel approach to defending against this relatively new threat.
\system uses honey connections to both deceive adversaries and dissuade them from performing attacks.
A key idea behind \system is to focus on client hosts performing high-value organizational roles, building metaphorical haystacks around their network traffic.
The honey connections used to build these haystacks also act as network canaries to bait adversaries and more quickly detect their presence.
We built a prototype of \system in an SDN environment and modeled its operation using the PRISM probabilistic model checker.
In doing so, we found that \system reliably ranks compromised switches among the most suspicious while having only a small effect on network request completion time.
As such, we believe role-based network deception is a promising approach for defending against adversaries that have compromised network devices.
\end{techreport}

\begin{acks}
  This research was partially sponsored by the Army Research Office and was 
  accomplished under Grant Number W911NF-17-1-0370. The views
  and conclusions contained in this document are those of the authors
  and should not be interpreted as representing the official policies,
  either expressed or implied, of the Army Research Office or the U.S.
  Government. The U.S. Government is authorized to reproduce and
  distribute reprints for Government purposes notwithstanding any
  copyright notation herein.
\end{acks}

\bibliographystyle{ACM-Reference-Format}
\bibliography{anjum_sdn,anjum_other,muzhu_bib,enck}


\begin{thebibliography}{79}


\ifx \showCODEN    \undefined \def \showCODEN     #1{\unskip}     \fi
\ifx \showDOI      \undefined \def \showDOI       #1{#1}\fi
\ifx \showISBNx    \undefined \def \showISBNx     #1{\unskip}     \fi
\ifx \showISBNxiii \undefined \def \showISBNxiii  #1{\unskip}     \fi
\ifx \showISSN     \undefined \def \showISSN      #1{\unskip}     \fi
\ifx \showLCCN     \undefined \def \showLCCN      #1{\unskip}     \fi
\ifx \shownote     \undefined \def \shownote      #1{#1}          \fi
\ifx \showarticletitle \undefined \def \showarticletitle #1{#1}   \fi
\ifx \showURL      \undefined \def \showURL       {\relax}        \fi
\providecommand\bibfield[2]{#2}
\providecommand\bibinfo[2]{#2}
\providecommand\natexlab[1]{#1}
\providecommand\showeprint[2][]{arXiv:#2}

\bibitem[\protect\citeauthoryear{Achleitner, La~Porta, McDaniel, Sugrim,
  Krishnamurthy, and Chadha}{Achleitner et~al\mbox{.}}{2016}]%
        {Achleitner:CyberDeception:2016}
\bibfield{author}{\bibinfo{person}{Stefan Achleitner}, \bibinfo{person}{Thomas
  La~Porta}, \bibinfo{person}{Patrick McDaniel}, \bibinfo{person}{Shridatt
  Sugrim}, \bibinfo{person}{Srikanth~V. Krishnamurthy}, {and}
  \bibinfo{person}{Ritu Chadha}.} \bibinfo{year}{2016}\natexlab{}.
\newblock \showarticletitle{Cyber Deception: Virtual Networks to Defend Insider
  Reconnaissance}. In \bibinfo{booktitle}{\emph{Proceedings of the 8th ACM CCS
  International Workshop on Managing Insider Security Threats}}.
  \bibinfo{publisher}{ACM}, \bibinfo{address}{New York},
  \bibinfo{pages}{57--68}.
\newblock
\showISBNx{978-1-4503-4571-2}
\urldef\tempurl%
\url{https://doi.org/10.1145/2995959.2995962}
\showDOI{\tempurl}


\bibitem[\protect\citeauthoryear{Agha and Palmskog}{Agha and Palmskog}{2018}]%
        {SMC}
\bibfield{author}{\bibinfo{person}{Gul Agha} {and} \bibinfo{person}{Karl
  Palmskog}.} \bibinfo{year}{2018}\natexlab{}.
\newblock \showarticletitle{A Survey of Statistical Model Checking}.
\newblock \bibinfo{journal}{\emph{ACM Trans. Model. Comput. Simul.}}
  \bibinfo{volume}{28}, \bibinfo{number}{1}, Article
  \bibinfo{articleno}{Article 6} (\bibinfo{date}{Jan.} \bibinfo{year}{2018}),
  \bibinfo{numpages}{39}~pages.
\newblock
\showISSN{1049-3301}
\urldef\tempurl%
\url{https://doi.org/10.1145/3158668}
\showDOI{\tempurl}


\bibitem[\protect\citeauthoryear{Almeshekah and Spafford}{Almeshekah and
  Spafford}{2016}]%
        {almeshekah2016}
\bibfield{author}{\bibinfo{person}{Mohammed~H Almeshekah} {and}
  \bibinfo{person}{Eugene~H Spafford}.} \bibinfo{year}{2016}\natexlab{}.
\newblock \showarticletitle{Cyber security deception}.
\newblock In \bibinfo{booktitle}{\emph{Cyber deception}}.
  \bibinfo{publisher}{Springer}, \bibinfo{pages}{23--50}.
\newblock


\bibitem[\protect\citeauthoryear{Anderson and McGrew}{Anderson and
  McGrew}{2019}]%
        {TSL:2019}
\bibfield{author}{\bibinfo{person}{Blake Anderson} {and} \bibinfo{person}{David
  McGrew}.} \bibinfo{year}{2019}\natexlab{}.
\newblock \showarticletitle{TLS Beyond the Browser: Combining End Host and
  Network Data to Understand Application Behavior}. In
  \bibinfo{booktitle}{\emph{Proceedings of the Internet Measurement
  Conference}} \emph{(\bibinfo{series}{IMC ’19})}.
  \bibinfo{publisher}{Association for Computing Machinery},
  \bibinfo{address}{New York, NY, USA}, \bibinfo{pages}{379–392}.
\newblock
\showISBNx{9781450369480}
\urldef\tempurl%
\url{https://doi.org/10.1145/3355369.3355601}
\showDOI{\tempurl}


\bibitem[\protect\citeauthoryear{Anjum, Miah, Zhu, Sharmin, Kiekintveld, Enck,
  and Singh}{Anjum et~al\mbox{.}}{2020}]%
        {anjum2020optimizing}
\bibfield{author}{\bibinfo{person}{Iffat Anjum},
  \bibinfo{person}{Mohammad~Sujan Miah}, \bibinfo{person}{Mu Zhu},
  \bibinfo{person}{Nazia Sharmin}, \bibinfo{person}{Christopher Kiekintveld},
  \bibinfo{person}{William Enck}, {and} \bibinfo{person}{Munindar~P Singh}.}
  \bibinfo{year}{2020}\natexlab{}.
\newblock \bibinfo{title}{Optimizing Vulnerability-Driven Honey Traffic Using
  Game Theory}.
\newblock
\newblock
\showeprint[arxiv]{cs.CR/2002.09069}


\bibitem[\protect\citeauthoryear{Antikainen, Aura, and
  S{\"a}rel{\"a}}{Antikainen et~al\mbox{.}}{2014}]%
        {Markku:2014:Spook}
\bibfield{author}{\bibinfo{person}{Markku Antikainen}, \bibinfo{person}{Tuomas
  Aura}, {and} \bibinfo{person}{Mikko S{\"a}rel{\"a}}.}
  \bibinfo{year}{2014}\natexlab{}.
\newblock \showarticletitle{Spook in Your Network: Attacking an SDN with a
  Compromised OpenFlow Switch}. In \bibinfo{booktitle}{\emph{Secure IT
  Systems}}, \bibfield{editor}{\bibinfo{person}{Karin Bernsmed} {and}
  \bibinfo{person}{Simone Fischer-H{\"u}bner}} (Eds.).
  \bibinfo{publisher}{Springer International Publishing},
  \bibinfo{address}{Cham}, \bibinfo{pages}{229--244}.
\newblock
\showISBNx{978-3-319-11599-3}


\bibitem[\protect\citeauthoryear{Backes, Doychev, and Köpf}{Backes
  et~al\mbox{.}}{2013}]%
        {side}
\bibfield{author}{\bibinfo{person}{Michael Backes}, \bibinfo{person}{Goran
  Doychev}, {and} \bibinfo{person}{Boris Köpf}.}
  \bibinfo{year}{2013}\natexlab{}.
\newblock \showarticletitle{Preventing Side-Channel Leaks in Web Traffic: A
  Formal Approach}. In \bibinfo{booktitle}{\emph{20\textsuperscript{th} ISOC
  Network and Distributed System Security Symposium}}.
\newblock


\bibitem[\protect\citeauthoryear{Bartlett, Heidemann, and
  Papadopoulos}{Bartlett et~al\mbox{.}}{2007}]%
        {bartlett2007understanding}
\bibfield{author}{\bibinfo{person}{Genevieve Bartlett}, \bibinfo{person}{John
  Heidemann}, {and} \bibinfo{person}{Christos Papadopoulos}.}
  \bibinfo{year}{2007}\natexlab{}.
\newblock \showarticletitle{Understanding passive and active service
  discovery}. In \bibinfo{booktitle}{\emph{Proceedings of the 7th ACM SIGCOMM
  conference on Internet measurement}}. \bibinfo{pages}{57--70}.
\newblock


\bibitem[\protect\citeauthoryear{Benton, Camp, and Small}{Benton
  et~al\mbox{.}}{2013}]%
        {Benton:2013:OVA}
\bibfield{author}{\bibinfo{person}{Kevin Benton}, \bibinfo{person}{L.~Jean
  Camp}, {and} \bibinfo{person}{Chris Small}.} \bibinfo{year}{2013}\natexlab{}.
\newblock \showarticletitle{OpenFlow Vulnerability Assessment}. In
  \bibinfo{booktitle}{\emph{Proceedings of the Second ACM SIGCOMM Workshop on
  Hot Topics in Software Defined Networking}} \emph{(\bibinfo{series}{HotSDN
  '13})}. \bibinfo{publisher}{ACM}, \bibinfo{address}{New York, NY, USA},
  \bibinfo{pages}{151--152}.
\newblock
\showISBNx{978-1-4503-2178-5}
\urldef\tempurl%
\url{https://doi.org/10.1145/2491185.2491222}
\showDOI{\tempurl}


\bibitem[\protect\citeauthoryear{Bowen, Kemerlis, Prabhu, Keromytis, and
  Stolfo}{Bowen et~al\mbox{.}}{2012a}]%
        {Bowen:2012:SGI}
\bibfield{author}{\bibinfo{person}{Brian~M. Bowen},
  \bibinfo{person}{Vasileios~P. Kemerlis}, \bibinfo{person}{Pratap Prabhu},
  \bibinfo{person}{Angelos~D. Keromytis}, {and} \bibinfo{person}{Salvatore~J.
  Stolfo}.} \bibinfo{year}{2012}\natexlab{a}.
\newblock \showarticletitle{A System for Generating and Injecting
  Indistinguishable Network Decoys}.
\newblock \bibinfo{journal}{\emph{J. Comput. Secur.}} \bibinfo{volume}{20},
  \bibinfo{number}{2-3} (\bibinfo{year}{2012}), \bibinfo{pages}{199--221}.
\newblock


\bibitem[\protect\citeauthoryear{Bowen, Kemerlis, Prabhu, Keromytis, and
  Stolfo}{Bowen et~al\mbox{.}}{2012b}]%
        {decoy2012}
\bibfield{author}{\bibinfo{person}{Brian~M. Bowen},
  \bibinfo{person}{Vasileios~P. Kemerlis}, \bibinfo{person}{Pratap Prabhu},
  \bibinfo{person}{Angelos~D. Keromytis}, {and} \bibinfo{person}{Salvatore~J.
  Stolfo}.} \bibinfo{year}{2012}\natexlab{b}.
\newblock \showarticletitle{A System for Generating and Injecting
  Indistinguishable Network Decoys}.
\newblock \bibinfo{journal}{\emph{J. Comput. Secur.}} \bibinfo{volume}{20},
  \bibinfo{number}{2–3} (\bibinfo{date}{March} \bibinfo{year}{2012}),
  \bibinfo{pages}{199–221}.
\newblock
\showISSN{0926-227X}


\bibitem[\protect\citeauthoryear{{Casado}, {Freedman}, {Pettit}, {Luo}, {Gude},
  {McKeown}, and {Shenker}}{{Casado} et~al\mbox{.}}{2009}]%
        {Casado:2009:Enterprise}
\bibfield{author}{\bibinfo{person}{M. {Casado}}, \bibinfo{person}{M.~J.
  {Freedman}}, \bibinfo{person}{J. {Pettit}}, \bibinfo{person}{J. {Luo}},
  \bibinfo{person}{N. {Gude}}, \bibinfo{person}{N. {McKeown}}, {and}
  \bibinfo{person}{S. {Shenker}}.} \bibinfo{year}{2009}\natexlab{}.
\newblock \showarticletitle{Rethinking Enterprise Network Control}.
\newblock \bibinfo{journal}{\emph{IEEE/ACM Transactions on Networking}}
  \bibinfo{volume}{17}, \bibinfo{number}{4} (\bibinfo{date}{Aug}
  \bibinfo{year}{2009}), \bibinfo{pages}{1270--1283}.
\newblock
\showISSN{1063-6692}
\urldef\tempurl%
\url{https://doi.org/10.1109/TNET.2009.2026415}
\showDOI{\tempurl}


\bibitem[\protect\citeauthoryear{Center}{Center}{1996}]%
        {internet2}
\bibfield{author}{\bibinfo{person}{Internet2 Network~Operations Center}.}
  \bibinfo{year}{1996}\natexlab{}.
\newblock \bibinfo{title}{Internet2}.
\newblock
\newblock
\urldef\tempurl%
\url{https://www.internet2.edu/}
\showURL{%
\tempurl}


\bibitem[\protect\citeauthoryear{{Chao}, {Ke}, {Chen}, {Chen}, {Hsieh}, {Lee},
  and {Hsiao}}{{Chao} et~al\mbox{.}}{2016}]%
        {Chao:2016:SDNData}
\bibfield{author}{\bibinfo{person}{T. {Chao}}, \bibinfo{person}{Y. {Ke}},
  \bibinfo{person}{B. {Chen}}, \bibinfo{person}{J. {Chen}},
  \bibinfo{person}{C.~J. {Hsieh}}, \bibinfo{person}{S. {Lee}}, {and}
  \bibinfo{person}{H. {Hsiao}}.} \bibinfo{year}{2016}\natexlab{}.
\newblock \showarticletitle{Securing data planes in software-defined networks}.
  In \bibinfo{booktitle}{\emph{2016 IEEE NetSoft Conference and Workshops
  (NetSoft)}}. \bibinfo{pages}{465--470}.
\newblock
\urldef\tempurl%
\url{https://doi.org/10.1109/NETSOFT.2016.7502486}
\showDOI{\tempurl}


\bibitem[\protect\citeauthoryear{Chen, Desmet, and Huygens}{Chen
  et~al\mbox{.}}{2014}]%
        {chen2014study}
\bibfield{author}{\bibinfo{person}{Ping Chen}, \bibinfo{person}{Lieven Desmet},
  {and} \bibinfo{person}{Christophe Huygens}.} \bibinfo{year}{2014}\natexlab{}.
\newblock \showarticletitle{A study on advanced persistent threats}. In
  \bibinfo{booktitle}{\emph{IFIP International Conference on Communications and
  Multimedia Security}}. Springer, \bibinfo{pages}{63--72}.
\newblock


\bibitem[\protect\citeauthoryear{Chi, Kuo, Guo, and Lei}{Chi
  et~al\mbox{.}}{2015}]%
        {detectCompromisedSwitch:2015}
\bibfield{author}{\bibinfo{person}{Po-Wen Chi}, \bibinfo{person}{Chien-Ting
  Kuo}, \bibinfo{person}{Jing-Wei Guo}, {and} \bibinfo{person}{Chin-Laung
  Lei}.} \bibinfo{year}{2015}\natexlab{}.
\newblock \showarticletitle{How to detect a compromised {SDN} switch}. In
  \bibinfo{booktitle}{\emph{Proceedings of the 1st IEEE Conference on Network
  Softwarization (NetSoft)}}. \bibinfo{pages}{1--6}.
\newblock
\urldef\tempurl%
\url{https://doi.org/10.1109/NETSOFT.2015.7116184}
\showDOI{\tempurl}


\bibitem[\protect\citeauthoryear{{Chowdhury}, {Bari}, {Ahmed}, and
  {Boutaba}}{{Chowdhury} et~al\mbox{.}}{2014}]%
        {PayLess:2014:Chowdhury}
\bibfield{author}{\bibinfo{person}{S.~R. {Chowdhury}}, \bibinfo{person}{M.~F.
  {Bari}}, \bibinfo{person}{R. {Ahmed}}, {and} \bibinfo{person}{R. {Boutaba}}.}
  \bibinfo{year}{2014}\natexlab{}.
\newblock \showarticletitle{PayLess: A low cost network monitoring framework
  for Software Defined Networks}. In \bibinfo{booktitle}{\emph{2014 IEEE
  Network Operations and Management Symposium (NOMS)}}. \bibinfo{pages}{1--9}.
\newblock
\showISSN{1542-1201}
\urldef\tempurl%
\url{https://doi.org/10.1109/NOMS.2014.6838227}
\showDOI{\tempurl}


\bibitem[\protect\citeauthoryear{Cimpanu}{Cimpanu}{2019}]%
        {cim19}
\bibfield{author}{\bibinfo{person}{Catalin Cimpanu}.}
  \bibinfo{year}{2019}\natexlab{}.
\newblock \bibinfo{title}{{Cisco bungled RV320/RV325 patches, routers still
  exposed to hacks}}.
\newblock \bibinfo{howpublished}{ZDNet}.
\newblock
\newblock
\shownote{\url{https://www.zdnet.com/article/cisco-bungled-rv320rv325-patches-routers-still-exposed-to-hacks/}.}


\bibitem[\protect\citeauthoryear{Clark, Sun, and Poovendran}{Clark
  et~al\mbox{.}}{2013}]%
        {decoyIP}
\bibfield{author}{\bibinfo{person}{Andrew Clark}, \bibinfo{person}{Kun Sun},
  {and} \bibinfo{person}{Radha Poovendran}.} \bibinfo{year}{2013}\natexlab{}.
\newblock \showarticletitle{Effectiveness of IP address randomization in
  decoy-based moving target defense}. In
  \bibinfo{booktitle}{\emph{52\textsuperscript{nd} IEEE Conference on Decision
  and Control}}. \bibinfo{pages}{678--685}.
\newblock
\showISBNx{978-1-4673-5717-3}
\urldef\tempurl%
\url{https://doi.org/10.1109/CDC.2013.6759960}
\showDOI{\tempurl}


\bibitem[\protect\citeauthoryear{ClusterDesign.org}{ClusterDesign.org}{2013}]%
        {fatTree}
\bibfield{author}{\bibinfo{person}{ClusterDesign.org}.}
  \bibinfo{year}{2013}\natexlab{}.
\newblock \bibinfo{title}{Fat-Tree Design}.
\newblock
\newblock
\urldef\tempurl%
\url{https://clusterdesign.org/fat-trees/}
\showURL{%
\tempurl}


\bibitem[\protect\citeauthoryear{Cohen}{Cohen}{1988}]%
        {cohen:1988}
\bibfield{author}{\bibinfo{person}{Jacob Cohen}.}
  \bibinfo{year}{1988}\natexlab{}.
\newblock \bibinfo{booktitle}{\emph{Statistical Power Analysis for the
  Behavioral Sciences} (\bibinfo{edition}{second} ed.)}.
\newblock \bibinfo{publisher}{Routledge Member of the Taylor and Francis
  Group}.
\newblock


\bibitem[\protect\citeauthoryear{Conde}{Conde}{2017}]%
        {SDP}
\bibfield{author}{\bibinfo{person}{Daniel Conde}.}
  \bibinfo{year}{2017}\natexlab{}.
\newblock \bibinfo{title}{Software-Defined Perimeters: An Architectural View of
  SDP}.
\newblock
  \bibinfo{howpublished}{\url{https://sdn.ieee.org/newsletter/march-2017/software-defined-perimeters-an-architectural-view-of-sdp}}.
\newblock


\bibitem[\protect\citeauthoryear{Conti, De~Gaspari, and Mancini}{Conti
  et~al\mbox{.}}{2017}]%
        {Conti_2017}
\bibfield{author}{\bibinfo{person}{Mauro Conti}, \bibinfo{person}{Fabio
  De~Gaspari}, {and} \bibinfo{person}{Luigi~V. Mancini}.}
  \bibinfo{year}{2017}\natexlab{}.
\newblock \showarticletitle{Know Your Enemy: Stealth Configuration-Information
  Gathering in SDN}.
\newblock \bibinfo{journal}{\emph{Lecture Notes in Computer Science}}
  (\bibinfo{year}{2017}), \bibinfo{pages}{386–401}.
\newblock
\showISBNx{9783319571867}
\showISSN{1611-3349}
\urldef\tempurl%
\url{https://doi.org/10.1007/978-3-319-57186-7_29}
\showDOI{\tempurl}


\bibitem[\protect\citeauthoryear{Cox}{Cox}{2019}]%
        {cox19}
\bibfield{author}{\bibinfo{person}{Joseph Cox}.}
  \bibinfo{year}{2019}\natexlab{}.
\newblock \bibinfo{title}{{As Phones Get Harder to Hack, Zero Day Vendors Hunt
  for Router Exploits}}.
\newblock \bibinfo{howpublished}{Motherboard: Tech by Vice}.
\newblock
\newblock
\shownote{\url{https://www.vice.com/en_us/article/evek9z/phones-harder-to-hack-crowdfense-zerodium-buy-router-zero-days-exploits}.}


\bibitem[\protect\citeauthoryear{Dhawan, Poddar, Mahajan, and Mann}{Dhawan
  et~al\mbox{.}}{2015}]%
        {dhawan:2015:sphinx}
\bibfield{author}{\bibinfo{person}{Mohan Dhawan}, \bibinfo{person}{Rishabh
  Poddar}, \bibinfo{person}{Kshiteej Mahajan}, {and} \bibinfo{person}{Vijay
  Mann}.} \bibinfo{year}{2015}\natexlab{}.
\newblock \showarticletitle{SPHINX: Detecting Security Attacks in
  Software-Defined Networks}. In \bibinfo{booktitle}{\emph{ISOC Network and
  Distributed System Security Symposium}}.
\newblock
\urldef\tempurl%
\url{https://doi.org/10.14722/ndss.2015.23064}
\showDOI{\tempurl}


\bibitem[\protect\citeauthoryear{Enbody and Sood}{Enbody and Sood}{2013}]%
        {Enbody:2013:Targeted}
\bibfield{author}{\bibinfo{person}{R.~J. Enbody} {and} \bibinfo{person}{A.~K.
  Sood}.} \bibinfo{year}{2013}\natexlab{}.
\newblock \showarticletitle{Targeted Cyberattacks: A Superset of Advanced
  Persistent Threats}.
\newblock \bibinfo{journal}{\emph{IEEE Security \& Privacy}}
  \bibinfo{volume}{11}, \bibinfo{number}{01} (\bibinfo{year}{2013}),
  \bibinfo{pages}{54--61}.
\newblock
\showISSN{1540-7993}
\urldef\tempurl%
\url{https://doi.org/10.1109/MSP.2012.90}
\showDOI{\tempurl}


\bibitem[\protect\citeauthoryear{Fonseca, Porter, Katz, Shenker, and
  Stoica}{Fonseca et~al\mbox{.}}{2007}]%
        {Fonseca:2007:XPN}
\bibfield{author}{\bibinfo{person}{Rodrigo Fonseca}, \bibinfo{person}{George
  Porter}, \bibinfo{person}{Randy~H. Katz}, \bibinfo{person}{Scott Shenker},
  {and} \bibinfo{person}{Ion Stoica}.} \bibinfo{year}{2007}\natexlab{}.
\newblock \showarticletitle{X-trace: A Pervasive Network Tracing Framework}. In
  \bibinfo{booktitle}{\emph{Proceedings of the 4th USENIX Conference on
  Networked Systems Design \&\#38; Implementation}}
  \emph{(\bibinfo{series}{NSDI'07})}. \bibinfo{publisher}{USENIX Association},
  \bibinfo{address}{Berkeley, CA, USA}, \bibinfo{pages}{20--20}.
\newblock


\bibitem[\protect\citeauthoryear{Foundation}{Foundation}{1998}]%
        {Wireshark}
\bibfield{author}{\bibinfo{person}{Wireshark Foundation}.}
  \bibinfo{year}{1998}\natexlab{}.
\newblock \bibinfo{title}{Wireshark}.
\newblock \bibinfo{howpublished}{\url{https://www.wireshark.org/}}.
\newblock


\bibitem[\protect\citeauthoryear{Goutam, Enck, and Reaves}{Goutam
  et~al\mbox{.}}{2019}]%
        {Goutam:2019:HSL}
\bibfield{author}{\bibinfo{person}{Sanket Goutam}, \bibinfo{person}{William
  Enck}, {and} \bibinfo{person}{Bradley Reaves}.}
  \bibinfo{year}{2019}\natexlab{}.
\newblock \showarticletitle{Hestia: Simple Least Privilege Network Policies for
  Smart Homes}. In \bibinfo{booktitle}{\emph{Proceedings of the 12th Conference
  on Security and Privacy in Wireless and Mobile Networks}}
  \emph{(\bibinfo{series}{WiSec '19})}. \bibinfo{publisher}{ACM},
  \bibinfo{address}{New York, NY, USA}, \bibinfo{pages}{215--220}.
\newblock
\showISBNx{978-1-4503-6726-4}


\bibitem[\protect\citeauthoryear{Han, Zhao, Doup{\'e}, and Ahn}{Han
  et~al\mbox{.}}{2016}]%
        {han2016honeymix}
\bibfield{author}{\bibinfo{person}{Wonkyu Han}, \bibinfo{person}{Ziming Zhao},
  \bibinfo{person}{Adam Doup{\'e}}, {and} \bibinfo{person}{Gail-Joon Ahn}.}
  \bibinfo{year}{2016}\natexlab{}.
\newblock \showarticletitle{Honeymix: Toward sdn-based intelligent honeynet}.
  In \bibinfo{booktitle}{\emph{Proceedings of the 2016 ACM International
  Workshop on Security in Software Defined Networks \& Network Function
  Virtualization}}. ACM, \bibinfo{pages}{1--6}.
\newblock


\bibitem[\protect\citeauthoryear{Holmes}{Holmes}{2015}]%
        {SYNfulKnock:2015}
\bibfield{author}{\bibinfo{person}{Graham Holmes}.}
  \bibinfo{year}{2015}\natexlab{}.
\newblock \bibinfo{title}{Evolution of attacks on Cisco IOS devices}.
\newblock
  \bibinfo{howpublished}{\url{https://blogs.cisco.com/security/evolution-of-attacks-on-cisco-ios-devices}}.
\newblock


\bibitem[\protect\citeauthoryear{Houmansadr, Brubaker, and
  Shmatikov}{Houmansadr et~al\mbox{.}}{2013}]%
        {ParrotDead}
\bibfield{author}{\bibinfo{person}{Amir Houmansadr}, \bibinfo{person}{Chad
  Brubaker}, {and} \bibinfo{person}{Vitaly Shmatikov}.}
  \bibinfo{year}{2013}\natexlab{}.
\newblock \showarticletitle{The parrot is dead: Observing unobservable network
  communications}. In \bibinfo{booktitle}{\emph{Proceedings - 2013 IEEE
  Symposium on Security and Privacy, SP 2013}}
  \emph{(\bibinfo{series}{Proceedings - IEEE Symposium on Security and
  Privacy})}. \bibinfo{pages}{65--79}.
\newblock
\showISBNx{9780769549774}
\urldef\tempurl%
\url{https://doi.org/10.1109/SP.2013.14}
\showDOI{\tempurl}


\bibitem[\protect\citeauthoryear{{Jafarian}, {Al-Shaer}, and {Duan}}{{Jafarian}
  et~al\mbox{.}}{2015}]%
        {Jafarian2015}
\bibfield{author}{\bibinfo{person}{J.~H. {Jafarian}}, \bibinfo{person}{E.
  {Al-Shaer}}, {and} \bibinfo{person}{Q. {Duan}}.}
  \bibinfo{year}{2015}\natexlab{}.
\newblock \showarticletitle{An Effective Address Mutation Approach for
  Disrupting Reconnaissance Attacks}.
\newblock \bibinfo{journal}{\emph{IEEE Transactions on Information Forensics
  and Security}} \bibinfo{volume}{10}, \bibinfo{number}{12}
  (\bibinfo{date}{Dec} \bibinfo{year}{2015}), \bibinfo{pages}{2562--2577}.
\newblock
\showISSN{1556-6021}
\urldef\tempurl%
\url{https://doi.org/10.1109/TIFS.2015.2467358}
\showDOI{\tempurl}


\bibitem[\protect\citeauthoryear{Kazemian, Varghese, and McKeown}{Kazemian
  et~al\mbox{.}}{2012}]%
        {HSA:2012}
\bibfield{author}{\bibinfo{person}{Peyman Kazemian}, \bibinfo{person}{George
  Varghese}, {and} \bibinfo{person}{Nick McKeown}.}
  \bibinfo{year}{2012}\natexlab{}.
\newblock \showarticletitle{Header Space Analysis: Static Checking for
  Networks}. In \bibinfo{booktitle}{\emph{Proceedings of the 9th USENIX
  Conference on Networked Systems Design and Implementation}}
  \emph{(\bibinfo{series}{NSDI’12})}. \bibinfo{publisher}{USENIX
  Association}, \bibinfo{address}{USA}, \bibinfo{pages}{9}.
\newblock


\bibitem[\protect\citeauthoryear{Khurshid, Zou, Zhou, Caesar, and
  Godfrey}{Khurshid et~al\mbox{.}}{2013}]%
        {VeriFlow:2013}
\bibfield{author}{\bibinfo{person}{Ahmed Khurshid}, \bibinfo{person}{Xuan Zou},
  \bibinfo{person}{Wenxuan Zhou}, \bibinfo{person}{Matthew Caesar}, {and}
  \bibinfo{person}{P.~Brighten Godfrey}.} \bibinfo{year}{2013}\natexlab{}.
\newblock \showarticletitle{VeriFlow: Verifying Network-Wide Invariants in Real
  Time}. In \bibinfo{booktitle}{\emph{Presented as part of the 10th {USENIX}
  Symposium on Networked Systems Design and Implementation ({NSDI} 13)}}.
  \bibinfo{publisher}{{USENIX}}, \bibinfo{address}{Lombard, IL},
  \bibinfo{pages}{15--27}.
\newblock
\showISBNx{978-1-931971-00-3}


\bibitem[\protect\citeauthoryear{Kim, Basescu, Jia, Lee, Hu, and Perrig}{Kim
  et~al\mbox{.}}{2014}]%
        {Lightweight:Kim:2014}
\bibfield{author}{\bibinfo{person}{Tiffany Hyun-Jin Kim},
  \bibinfo{person}{Cristina Basescu}, \bibinfo{person}{Limin Jia},
  \bibinfo{person}{Soo~Bum Lee}, \bibinfo{person}{Yih-Chun Hu}, {and}
  \bibinfo{person}{Adrian Perrig}.} \bibinfo{year}{2014}\natexlab{}.
\newblock \showarticletitle{Lightweight Source Authentication and Path
  Validation}. In \bibinfo{booktitle}{\emph{Proceedings of the ACM Conference
  on SIGCOMM}}. \bibinfo{publisher}{ACM}, \bibinfo{address}{New York, NY, USA},
  \bibinfo{pages}{271--282}.
\newblock
\urldef\tempurl%
\url{https://doi.org/10.1145/2619239.2626323}
\showDOI{\tempurl}


\bibitem[\protect\citeauthoryear{Kranch and Bonneau}{Kranch and
  Bonneau}{2015}]%
        {Kranch:2015:UpgradingHTTPS}
\bibfield{author}{\bibinfo{person}{Michael Kranch} {and}
  \bibinfo{person}{Joseph Bonneau}.} \bibinfo{year}{2015}\natexlab{}.
\newblock \showarticletitle{Upgrading HTTPS in Mid-Air: An Empirical Study of
  Strict Transport Security and Key Pinning}. In \bibinfo{booktitle}{\emph{22nd
  Network and Distributed System Security Symposium {NDSS}}}.
\newblock


\bibitem[\protect\citeauthoryear{Krombholz, Mayer, Schmiedecker, and
  Weippl}{Krombholz et~al\mbox{.}}{2017}]%
        {Krombholz:2017:HTTPS}
\bibfield{author}{\bibinfo{person}{Katharina Krombholz},
  \bibinfo{person}{Wilfired Mayer}, \bibinfo{person}{Martin Schmiedecker},
  {and} \bibinfo{person}{Edgar Weippl}.} \bibinfo{year}{2017}\natexlab{}.
\newblock \showarticletitle{``I have No Idea What I'm Doing'' - On the
  Usability of Deploying HTTPS}. In \bibinfo{booktitle}{\emph{26th {USENIX}
  Security Symposium ({USENIX} Security 17)}}. \bibinfo{publisher}{{USENIX}
  Association}.
\newblock
\showISBNx{978-1-931971-40-9}


\bibitem[\protect\citeauthoryear{Kwiatkowska, Norman, and Parker}{Kwiatkowska
  et~al\mbox{.}}{2011}]%
        {PRISM}
\bibfield{author}{\bibinfo{person}{M. Kwiatkowska}, \bibinfo{person}{G.
  Norman}, {and} \bibinfo{person}{D. Parker}.} \bibinfo{year}{2011}\natexlab{}.
\newblock \showarticletitle{{PRISM} 4.0: Verification of Probabilistic
  Real-time Systems}. In \bibinfo{booktitle}{\emph{Proc. 23rd International
  Conference on Computer Aided Verification (CAV'11)}}
  \emph{(\bibinfo{series}{LNCS})},
  \bibfield{editor}{\bibinfo{person}{G.~Gopalakrishnan} {and}
  \bibinfo{person}{S.~Qadeer}} (Eds.), Vol.~\bibinfo{volume}{6806}.
  \bibinfo{publisher}{Springer}, \bibinfo{pages}{585--591}.
\newblock


\bibitem[\protect\citeauthoryear{Kwiatkowska, Norman, and Parker}{Kwiatkowska
  et~al\mbox{.}}{2018}]%
        {Kwiatkowska2018}
\bibfield{author}{\bibinfo{person}{Marta Kwiatkowska}, \bibinfo{person}{Gethin
  Norman}, {and} \bibinfo{person}{David Parker}.}
  \bibinfo{year}{2018}\natexlab{}.
\newblock \bibinfo{booktitle}{\emph{Probabilistic Model Checking: Advances and
  Applications}}.
\newblock \bibinfo{publisher}{Springer International Publishing},
  \bibinfo{address}{Cham}, \bibinfo{pages}{73--121}.
\newblock


\bibitem[\protect\citeauthoryear{Levin, Canini, Schmid, Schaffert, and
  Feldmann}{Levin et~al\mbox{.}}{2014}]%
        {Levin:2014:Panopticon}
\bibfield{author}{\bibinfo{person}{Dan Levin}, \bibinfo{person}{Marco Canini},
  \bibinfo{person}{Stefan Schmid}, \bibinfo{person}{Fabian Schaffert}, {and}
  \bibinfo{person}{Anja Feldmann}.} \bibinfo{year}{2014}\natexlab{}.
\newblock \showarticletitle{Panopticon: Reaping the Benefits of Incremental
  {SDN} Deployment in Enterprise Networks}. In \bibinfo{booktitle}{\emph{2014
  {USENIX} Annual Technical Conference ({USENIX} {ATC} 14)}}.
  \bibinfo{publisher}{{USENIX} Association}, \bibinfo{address}{Philadelphia,
  PA}, \bibinfo{pages}{333--345}.
\newblock
\showISBNx{978-1-931971-10-2}


\bibitem[\protect\citeauthoryear{{Li}, {Lai}, and {Ddl}}{{Li}
  et~al\mbox{.}}{2011}]%
        {Li:2011:APT}
\bibfield{author}{\bibinfo{person}{F. {Li}}, \bibinfo{person}{A. {Lai}}, {and}
  \bibinfo{person}{D. {Ddl}}.} \bibinfo{year}{2011}\natexlab{}.
\newblock \showarticletitle{Evidence of Advanced Persistent Threat: A case
  study of malware for political espionage}. In \bibinfo{booktitle}{\emph{2011
  6th International Conference on Malicious and Unwanted Software}}.
  \bibinfo{pages}{102--109}.
\newblock
\urldef\tempurl%
\url{https://doi.org/10.1109/MALWARE.2011.6112333}
\showDOI{\tempurl}


\bibitem[\protect\citeauthoryear{{Li}, {Zou}, {Huang}, {Zheng}, and {Lee}}{{Li}
  et~al\mbox{.}}{2018}]%
        {Li:2018:Dyna}
\bibfield{author}{\bibinfo{person}{Q. {Li}}, \bibinfo{person}{X. {Zou}},
  \bibinfo{person}{Q. {Huang}}, \bibinfo{person}{J. {Zheng}}, {and}
  \bibinfo{person}{P.~P.~C. {Lee}}.} \bibinfo{year}{2018}\natexlab{}.
\newblock \showarticletitle{Dynamic Packet Forwarding Verification in SDN}.
\newblock \bibinfo{journal}{\emph{IEEE Transactions on Dependable and Secure
  Computing}} (\bibinfo{year}{2018}), \bibinfo{pages}{1--1}.
\newblock
\showISSN{1545-5971}
\urldef\tempurl%
\url{https://doi.org/10.1109/TDSC.2018.2810880}
\showDOI{\tempurl}


\bibitem[\protect\citeauthoryear{{Ling}, {Luo}, {Zhang}, {Ming Yang}, {Fu}, and
  {Yu}}{{Ling} et~al\mbox{.}}{2012}]%
        {delay}
\bibfield{author}{\bibinfo{person}{Z. {Ling}}, \bibinfo{person}{J. {Luo}},
  \bibinfo{person}{Y. {Zhang}}, \bibinfo{person}{{Ming Yang}},
  \bibinfo{person}{X. {Fu}}, {and} \bibinfo{person}{W. {Yu}}.}
  \bibinfo{year}{2012}\natexlab{}.
\newblock \showarticletitle{A novel network delay based side-channel attack:
  Modeling and defense}. In \bibinfo{booktitle}{\emph{2012 Proceedings IEEE
  INFOCOM}}. \bibinfo{pages}{2390--2398}.
\newblock
\showISSN{0743-166X}
\urldef\tempurl%
\url{https://doi.org/10.1109/INFCOM.2012.6195628}
\showDOI{\tempurl}


\bibitem[\protect\citeauthoryear{{Lorenz}, {Hock}, {Scherer}, {Durner},
  {Kellerer}, {Gebert}, {Gray}, {Zinner}, and {Tran-Gia}}{{Lorenz}
  et~al\mbox{.}}{2017}]%
        {Lorenz:2017:SDN-Enterprise}
\bibfield{author}{\bibinfo{person}{C. {Lorenz}}, \bibinfo{person}{D. {Hock}},
  \bibinfo{person}{J. {Scherer}}, \bibinfo{person}{R. {Durner}},
  \bibinfo{person}{W. {Kellerer}}, \bibinfo{person}{S. {Gebert}},
  \bibinfo{person}{N. {Gray}}, \bibinfo{person}{T. {Zinner}}, {and}
  \bibinfo{person}{P. {Tran-Gia}}.} \bibinfo{year}{2017}\natexlab{}.
\newblock \showarticletitle{An SDN/NFV-Enabled Enterprise Network Architecture
  Offering Fine-Grained Security Policy Enforcement}.
\newblock \bibinfo{journal}{\emph{IEEE Communications Magazine}}
  \bibinfo{volume}{55}, \bibinfo{number}{3} (\bibinfo{date}{March}
  \bibinfo{year}{2017}), \bibinfo{pages}{217--223}.
\newblock
\showISSN{0163-6804}
\urldef\tempurl%
\url{https://doi.org/10.1109/MCOM.2017.1600414CM}
\showDOI{\tempurl}


\bibitem[\protect\citeauthoryear{Luo, Laperdrix, Honarmand, and
  Nikiforakis}{Luo et~al\mbox{.}}{2019}]%
        {Luo:2019:TimeDoesNotHeal}
\bibfield{author}{\bibinfo{person}{Meng Luo}, \bibinfo{person}{Pierre
  Laperdrix}, \bibinfo{person}{Nima Honarmand}, {and} \bibinfo{person}{Nick
  Nikiforakis}.} \bibinfo{year}{2019}\natexlab{}.
\newblock \showarticletitle{Time Does Not Heal All Wounds: A Longitudinal
  Analysis of Security-Mechanism Support in Mobile Browsers}. In
  \bibinfo{booktitle}{\emph{26th Network and Distributed System Security
  Symposium (NDSS)}}.
\newblock


\bibitem[\protect\citeauthoryear{Ma, Lei, Wang, Zhang, Xu, and Li}{Ma
  et~al\mbox{.}}{2016}]%
        {adaptingHop}
\bibfield{author}{\bibinfo{person}{Duohe Ma}, \bibinfo{person}{Cheng Lei},
  \bibinfo{person}{Liming Wang}, \bibinfo{person}{Hongqi Zhang},
  \bibinfo{person}{Zhen Xu}, {and} \bibinfo{person}{Meng Li}.}
  \bibinfo{year}{2016}\natexlab{}.
\newblock \showarticletitle{A Self-adaptive Hopping Approach of Moving Target
  Defense to thwart Scanning Attacks}. In \bibinfo{booktitle}{\emph{Information
  and Communications Security}}, \bibfield{editor}{\bibinfo{person}{Kwok-Yan
  Lam}, \bibinfo{person}{Chi-Hung Chi}, {and} \bibinfo{person}{Sihan Qing}}
  (Eds.). \bibinfo{publisher}{Springer International Publishing},
  \bibinfo{address}{Cham}, \bibinfo{pages}{39--53}.
\newblock
\showISBNx{978-3-319-50011-9}


\bibitem[\protect\citeauthoryear{McKeown, Anderson, Balakrishnan, Parulkar,
  Peterson, Rexford, Shenker, and Turner}{McKeown et~al\mbox{.}}{2008}]%
        {McKeown:2008:OpenFlow}
\bibfield{author}{\bibinfo{person}{Nick McKeown}, \bibinfo{person}{Tom
  Anderson}, \bibinfo{person}{Hari Balakrishnan}, \bibinfo{person}{Guru
  Parulkar}, \bibinfo{person}{Larry Peterson}, \bibinfo{person}{Jennifer
  Rexford}, \bibinfo{person}{Scott Shenker}, {and} \bibinfo{person}{Jonathan
  Turner}.} \bibinfo{year}{2008}\natexlab{}.
\newblock \showarticletitle{OpenFlow: Enabling Innovation in Campus Networks}.
\newblock \bibinfo{journal}{\emph{SIGCOMM Comput. Commun. Rev.}}
  \bibinfo{volume}{38}, \bibinfo{number}{2} (\bibinfo{year}{2008}),
  \bibinfo{pages}{69--74}.
\newblock
\showISSN{0146-4833}
\urldef\tempurl%
\url{https://doi.org/10.1145/1355734.1355746}
\showDOI{\tempurl}


\bibitem[\protect\citeauthoryear{Miessler}{Miessler}{2019}]%
        {amass:2019}
\bibfield{author}{\bibinfo{person}{Daniel Miessler}.}
  \bibinfo{year}{2019}\natexlab{}.
\newblock \bibinfo{title}{amass- Automated Attack Surface Mapping}.
\newblock
  \bibinfo{howpublished}{\url{https://danielmiessler.com/study/amass/}}.
\newblock


\bibitem[\protect\citeauthoryear{Miskovic, Lee, Liao, and Baldi}{Miskovic
  et~al\mbox{.}}{2015}]%
        {miskovic2015appprint}
\bibfield{author}{\bibinfo{person}{Stanislav Miskovic},
  \bibinfo{person}{Gene~Moo Lee}, \bibinfo{person}{Yong Liao}, {and}
  \bibinfo{person}{Mario Baldi}.} \bibinfo{year}{2015}\natexlab{}.
\newblock \showarticletitle{Appprint: automatic fingerprinting of mobile
  applications in network traffic}. In \bibinfo{booktitle}{\emph{International
  Conference on Passive and Active Network Measurement}}. Springer,
  \bibinfo{pages}{57--69}.
\newblock


\bibitem[\protect\citeauthoryear{Mizrak, Cheng, Marzullo, and Savage}{Mizrak
  et~al\mbox{.}}{2006}]%
        {MalRouter2}
\bibfield{author}{\bibinfo{person}{Alper~Tugay Mizrak},
  \bibinfo{person}{Yu-Chung Cheng}, \bibinfo{person}{Keith Marzullo}, {and}
  \bibinfo{person}{Stefan Savage}.} \bibinfo{year}{2006}\natexlab{}.
\newblock \showarticletitle{Detecting and Isolating Malicious Routers}.
\newblock \bibinfo{journal}{\emph{IEEE Trans. Dependable Secur. Comput.}}
  \bibinfo{volume}{3}, \bibinfo{number}{3} (\bibinfo{date}{July}
  \bibinfo{year}{2006}), \bibinfo{pages}{230–244}.
\newblock
\showISSN{1545-5971}
\urldef\tempurl%
\url{https://doi.org/10.1109/TDSC.2006.34}
\showDOI{\tempurl}


\bibitem[\protect\citeauthoryear{Mohamed, O'Connor, Miettinen, Enck, and
  Sadeghi}{Mohamed et~al\mbox{.}}{2019}]%
        {HoneyScope:2019}
\bibfield{author}{\bibinfo{person}{Reham Mohamed}, \bibinfo{person}{Terrance
  O'Connor}, \bibinfo{person}{Markus Miettinen}, \bibinfo{person}{William
  Enck}, {and} \bibinfo{person}{Ahmad-Reza Sadeghi}.}
  \bibinfo{year}{2019}\natexlab{}.
\newblock \showarticletitle{HONEYSCOPE: IoT Device Protection with Deceptive
  Network Views,}.
\newblock In \bibinfo{booktitle}{\emph{Autonomous Cyber Deception: Reasoning,
  Adaptive Planning, and Evaluation of HoneyThings}}.
  \bibinfo{publisher}{Springer International Publishing}.
\newblock


\bibitem[\protect\citeauthoryear{Nandikotkur}{Nandikotkur}{2018}]%
        {CiscoSwitch:2018}
\bibfield{author}{\bibinfo{person}{Geetha Nandikotkur}.}
  \bibinfo{year}{2018}\natexlab{}.
\newblock \bibinfo{title}{Evolution of attacks on Cisco IOS devices}.
\newblock
  \bibinfo{howpublished}{\url{https://www.bankinfosecurity.com/200000-cisco-network-switches-reportedly-hacked-a-10788}}.
\newblock


\bibitem[\protect\citeauthoryear{Nayak, Reimers, Feamster, and Clark}{Nayak
  et~al\mbox{.}}{2009}]%
        {Nayak:2009:RDA}
\bibfield{author}{\bibinfo{person}{Ankur~Kumar Nayak}, \bibinfo{person}{Alex
  Reimers}, \bibinfo{person}{Nick Feamster}, {and} \bibinfo{person}{Russ
  Clark}.} \bibinfo{year}{2009}\natexlab{}.
\newblock \showarticletitle{Resonance: Dynamic Access Control for Enterprise
  Networks}. In \bibinfo{booktitle}{\emph{Proceedings of the 1st ACM Workshop
  on Research on Enterprise Networking}} \emph{(\bibinfo{series}{WREN '09})}.
  \bibinfo{publisher}{ACM}, \bibinfo{address}{New York, NY, USA},
  \bibinfo{pages}{11--18}.
\newblock
\showISBNx{978-1-60558-443-0}
\urldef\tempurl%
\url{https://doi.org/10.1145/1592681.1592684}
\showDOI{\tempurl}


\bibitem[\protect\citeauthoryear{Networks}{Networks}{2019}]%
        {FlowMon}
\bibfield{author}{\bibinfo{person}{Flowmon Networks}.}
  \bibinfo{year}{2019}\natexlab{}.
\newblock \bibinfo{title}{Flowmon: Driving Network Visibility}.
\newblock \bibinfo{howpublished}{\url{https://www.flowmon.com/en/}}.
\newblock


\bibitem[\protect\citeauthoryear{OConnor, Enck, Petullo, and Verma}{OConnor
  et~al\mbox{.}}{2018}]%
        {OConnor:2018:PivotWall}
\bibfield{author}{\bibinfo{person}{Tj OConnor}, \bibinfo{person}{William Enck},
  \bibinfo{person}{W.~Michael Petullo}, {and} \bibinfo{person}{Akash Verma}.}
  \bibinfo{year}{2018}\natexlab{}.
\newblock \showarticletitle{PivotWall: SDN-Based Information Flow Control}. In
  \bibinfo{booktitle}{\emph{Proceedings of the Symposium on SDN Research}}
  \emph{(\bibinfo{series}{SOSR '18})}. \bibinfo{publisher}{ACM}, Article
  \bibinfo{articleno}{3}, \bibinfo{numpages}{14}~pages.
\newblock
\showISBNx{978-1-4503-5664-0}
\urldef\tempurl%
\url{https://doi.org/10.1145/3185467.3185474}
\showDOI{\tempurl}


\bibitem[\protect\citeauthoryear{of~Adelaide}{of~Adelaide}{2010}]%
        {TopoZoo}
\bibfield{author}{\bibinfo{person}{The~University of Adelaide}.}
  \bibinfo{year}{2010}\natexlab{}.
\newblock \bibinfo{title}{The Internet Topology Zoo}.
\newblock
\newblock
\urldef\tempurl%
\url{http://www.topology-zoo.org/contact.html}
\showURL{%
\tempurl}


\bibitem[\protect\citeauthoryear{Padmanabhan and Simon}{Padmanabhan and
  Simon}{2003}]%
        {MalRouter1}
\bibfield{author}{\bibinfo{person}{Venkata~N. Padmanabhan} {and}
  \bibinfo{person}{Daniel~R. Simon}.} \bibinfo{year}{2003}\natexlab{}.
\newblock \showarticletitle{Secure Traceroute to Detect Faulty or Malicious
  Routing}.
\newblock \bibinfo{journal}{\emph{SIGCOMM Comput. Commun. Rev.}}
  \bibinfo{volume}{33}, \bibinfo{number}{1} (\bibinfo{date}{Jan.}
  \bibinfo{year}{2003}), \bibinfo{pages}{77–82}.
\newblock
\showISSN{0146-4833}
\urldef\tempurl%
\url{https://doi.org/10.1145/774763.774775}
\showDOI{\tempurl}


\bibitem[\protect\citeauthoryear{Park, Woo, Moon, and Choi}{Park
  et~al\mbox{.}}{2018}]%
        {decoy2018}
\bibfield{author}{\bibinfo{person}{Kyungmin Park}, \bibinfo{person}{Samuel
  Woo}, \bibinfo{person}{Daesung Moon}, {and} \bibinfo{person}{Hoon Choi}.}
  \bibinfo{year}{2018}\natexlab{}.
\newblock \showarticletitle{Secure Cyber Deception Architecture and Decoy
  Injection to Mitigate the Insider Threat}.
\newblock \bibinfo{journal}{\emph{Symmetry}}  \bibinfo{volume}{10}
  (\bibinfo{date}{01} \bibinfo{year}{2018}), \bibinfo{pages}{14}.
\newblock
\urldef\tempurl%
\url{https://doi.org/10.3390/sym10010014}
\showDOI{\tempurl}


\bibitem[\protect\citeauthoryear{Qazi, Tu, Chiang, Miao, Sekar, and Yu}{Qazi
  et~al\mbox{.}}{2013}]%
        {Qazi:2013:SMP}
\bibfield{author}{\bibinfo{person}{Zafar~Ayyub Qazi},
  \bibinfo{person}{Cheng-Chun Tu}, \bibinfo{person}{Luis Chiang},
  \bibinfo{person}{Rui Miao}, \bibinfo{person}{Vyas Sekar}, {and}
  \bibinfo{person}{Minlan Yu}.} \bibinfo{year}{2013}\natexlab{}.
\newblock \showarticletitle{SIMPLE-fying Middlebox Policy Enforcement Using
  SDN}. In \bibinfo{booktitle}{\emph{Proceedings of the ACM SIGCOMM 2013
  Conference on SIGCOMM}} \emph{(\bibinfo{series}{SIGCOMM '13})}.
  \bibinfo{publisher}{ACM}, \bibinfo{pages}{27--38}.
\newblock
\showISBNx{978-1-4503-2056-6}
\urldef\tempurl%
\url{https://doi.org/10.1145/2486001.2486022}
\showDOI{\tempurl}


\bibitem[\protect\citeauthoryear{Rowe, Custy, and Duong}{Rowe
  et~al\mbox{.}}{2007}]%
        {Honeypots}
\bibfield{author}{\bibinfo{person}{Neil~C. Rowe}, \bibinfo{person}{E.~John
  Custy}, {and} \bibinfo{person}{Binh~T. Duong}.}
  \bibinfo{year}{2007}\natexlab{}.
\newblock \bibinfo{title}{Defending Cyberspace with Fake Honeypots}.
\newblock
\newblock
\urldef\tempurl%
\url{https://calhoun.nps.edu/handle/10945/36428}
\showURL{%
\tempurl}


\bibitem[\protect\citeauthoryear{Schuster, Shmatikov, and Tromer}{Schuster
  et~al\mbox{.}}{2017}]%
        {Beauty}
\bibfield{author}{\bibinfo{person}{Roei Schuster}, \bibinfo{person}{Vitaly
  Shmatikov}, {and} \bibinfo{person}{Eran Tromer}.}
  \bibinfo{year}{2017}\natexlab{}.
\newblock \showarticletitle{Beauty and the Burst: Remote Identification of
  Encrypted Video Streams}. In \bibinfo{booktitle}{\emph{26th {USENIX} Security
  Symposium ({USENIX} Security 17)}}. \bibinfo{publisher}{{USENIX}
  Association}, \bibinfo{address}{Vancouver, BC}, \bibinfo{pages}{1357--1374}.
\newblock
\showISBNx{978-1-931971-40-9}


\bibitem[\protect\citeauthoryear{Shaghaghi, Kaafar, and Jha}{Shaghaghi
  et~al\mbox{.}}{2017}]%
        {shaghaghi:Wedgetail:2017}
\bibfield{author}{\bibinfo{person}{Arash Shaghaghi},
  \bibinfo{person}{Mohamed~Ali Kaafar}, {and} \bibinfo{person}{Sanjay Jha}.}
  \bibinfo{year}{2017}\natexlab{}.
\newblock \showarticletitle{WedgeTail: An Intrusion Prevention System for the
  Data Plane of Software Defined Networks}. In
  \bibinfo{booktitle}{\emph{Proceedings of the 2017 ACM on Asia Conference on
  Computer and Communications Security}}. \bibinfo{publisher}{ACM},
  \bibinfo{address}{New York}, \bibinfo{pages}{849--861}.
\newblock
\showISBNx{978-1-4503-4944-4}
\urldef\tempurl%
\url{https://doi.org/10.1145/3052973.3053039}
\showDOI{\tempurl}


\bibitem[\protect\citeauthoryear{Snyder}{Snyder}{2014}]%
        {snowden:2014}
\bibfield{author}{\bibinfo{person}{Bill Snyder}.}
  \bibinfo{year}{2014}\natexlab{}.
\newblock \bibinfo{title}{Snowden: The NSA planted backdoors in Cisco
  products}.
\newblock
  \bibinfo{howpublished}{\url{https://www.infoworld.com/article/2608141/snowden--the-nsa-planted-backdoors-in-cisco-products.html}}.
\newblock


\bibitem[\protect\citeauthoryear{Sommers and Barford}{Sommers and
  Barford}{2004}]%
        {Sommers:2004:SNT}
\bibfield{author}{\bibinfo{person}{Joel Sommers} {and} \bibinfo{person}{Paul
  Barford}.} \bibinfo{year}{2004}\natexlab{}.
\newblock \showarticletitle{Self-configuring Network Traffic Generation}. In
  \bibinfo{booktitle}{\emph{Proceedings of the 4th ACM SIGCOMM Conference on
  Internet Measurement}} \emph{(\bibinfo{series}{IMC '04})}.
  \bibinfo{publisher}{ACM}, \bibinfo{address}{New York, NY, USA},
  \bibinfo{pages}{68--81}.
\newblock
\showISBNx{1-58113-821-0}
\urldef\tempurl%
\url{https://doi.org/10.1145/1028788.1028798}
\showDOI{\tempurl}


\bibitem[\protect\citeauthoryear{{Suh}, {Kwon}, {Dixon}, {Felter}, and
  {Carter}}{{Suh} et~al\mbox{.}}{2014}]%
        {OpenSample:2014:Suh}
\bibfield{author}{\bibinfo{person}{J. {Suh}}, \bibinfo{person}{T.~T. {Kwon}},
  \bibinfo{person}{C. {Dixon}}, \bibinfo{person}{W. {Felter}}, {and}
  \bibinfo{person}{J. {Carter}}.} \bibinfo{year}{2014}\natexlab{}.
\newblock \showarticletitle{OpenSample: A Low-Latency, Sampling-Based
  Measurement Platform for Commodity SDN}. In \bibinfo{booktitle}{\emph{2014
  IEEE 34th International Conference on Distributed Computing Systems}}.
  \bibinfo{pages}{228--237}.
\newblock
\showISSN{1063-6927}
\urldef\tempurl%
\url{https://doi.org/10.1109/ICDCS.2014.31}
\showDOI{\tempurl}


\bibitem[\protect\citeauthoryear{Target}{Target}{2019}]%
        {HT:2019}
\bibfield{author}{\bibinfo{person}{Hacker Target}.}
  \bibinfo{year}{2019}\natexlab{}.
\newblock \bibinfo{title}{Simplify the security assessment process with hosted
  vulnerability scanners}.
\newblock \bibinfo{howpublished}{\url{https://hackertarget.com/}}.
\newblock


\bibitem[\protect\citeauthoryear{Team}{Team}{2018}]%
        {Mininet}
\bibfield{author}{\bibinfo{person}{Mininet Team}.}
  \bibinfo{year}{2018}\natexlab{}.
\newblock \bibinfo{title}{Mininet An Instant Virtual Network on your Laptop (or
  other PC)}.
\newblock \bibinfo{howpublished}{\url{http://mininet.org/}}.
\newblock


\bibitem[\protect\citeauthoryear{Thimmaraju, Shastry, Fiebig, Hetzelt, Seifert,
  Feldmann, and Schmid}{Thimmaraju et~al\mbox{.}}{2016a}]%
        {Thimmaraju:Cloud:2016}
\bibfield{author}{\bibinfo{person}{Kashyap Thimmaraju},
  \bibinfo{person}{Bhargava Shastry}, \bibinfo{person}{Tobias Fiebig},
  \bibinfo{person}{Felicitas Hetzelt}, \bibinfo{person}{Jean{-}Pierre Seifert},
  \bibinfo{person}{Anja Feldmann}, {and} \bibinfo{person}{Stefan Schmid}.}
  \bibinfo{year}{2016}\natexlab{a}.
\newblock \showarticletitle{Reigns to the Cloud: Compromising Cloud Systems via
  the Data Plane}.
\newblock \bibinfo{journal}{\emph{CoRR}}  \bibinfo{volume}{abs/1610.08717}
  (\bibinfo{year}{2016}).
\newblock
\showeprint[arxiv]{1610.08717}


\bibitem[\protect\citeauthoryear{Thimmaraju, Shastry, Fiebig, Hetzelt, Seifert,
  Feldmann, and Schmid}{Thimmaraju et~al\mbox{.}}{2016b}]%
        {Thimmaraju:SFHSF:2016}
\bibfield{author}{\bibinfo{person}{Kashyap Thimmaraju},
  \bibinfo{person}{Bhargava Shastry}, \bibinfo{person}{Tobias Fiebig},
  \bibinfo{person}{Felicitas Hetzelt}, \bibinfo{person}{Jean{-}Pierre Seifert},
  \bibinfo{person}{Anja Feldmann}, {and} \bibinfo{person}{Stefan Schmid}.}
  \bibinfo{year}{2016}\natexlab{b}.
\newblock \showarticletitle{Reigns to the Cloud: Compromising Cloud Systems via
  the Data Plane}.
\newblock \bibinfo{journal}{\emph{CoRR}}  \bibinfo{volume}{abs/1610.08717}
  (\bibinfo{year}{2016}).
\newblock


\bibitem[\protect\citeauthoryear{{van Adrichem}, {Doerr}, and {Kuipers}}{{van
  Adrichem} et~al\mbox{.}}{2014}]%
        {OpenNetMon:2014:Adrichem}
\bibfield{author}{\bibinfo{person}{N.~L.~M. {van Adrichem}},
  \bibinfo{person}{C. {Doerr}}, {and} \bibinfo{person}{F.~A. {Kuipers}}.}
  \bibinfo{year}{2014}\natexlab{}.
\newblock \showarticletitle{OpenNetMon: Network monitoring in OpenFlow
  Software-Defined Networks}. In \bibinfo{booktitle}{\emph{2014 IEEE Network
  Operations and Management Symposium (NOMS)}}. \bibinfo{pages}{1--8}.
\newblock
\showISSN{1542-1201}
\urldef\tempurl%
\url{https://doi.org/10.1109/NOMS.2014.6838228}
\showDOI{\tempurl}


\bibitem[\protect\citeauthoryear{{Vishwanath} and {Vahdat}}{{Vishwanath} and
  {Vahdat}}{2009}]%
        {Swing:Vishwanath:2009}
\bibfield{author}{\bibinfo{person}{K.~V. {Vishwanath}} {and}
  \bibinfo{person}{A. {Vahdat}}.} \bibinfo{year}{2009}\natexlab{}.
\newblock \showarticletitle{Swing: Realistic and Responsive Network Traffic
  Generation}.
\newblock \bibinfo{journal}{\emph{IEEE/ACM Transactions on Networking}}
  \bibinfo{volume}{17}, \bibinfo{number}{3} (\bibinfo{date}{June}
  \bibinfo{year}{2009}), \bibinfo{pages}{712--725}.
\newblock
\showISSN{1063-6692}
\urldef\tempurl%
\url{https://doi.org/10.1109/TNET.2009.2020830}
\showDOI{\tempurl}


\bibitem[\protect\citeauthoryear{{Wang}, {Xu}, and {Gu}}{{Wang}
  et~al\mbox{.}}{2015}]%
        {Wang:2015:FloodGuard}
\bibfield{author}{\bibinfo{person}{H. {Wang}}, \bibinfo{person}{L. {Xu}}, {and}
  \bibinfo{person}{G. {Gu}}.} \bibinfo{year}{2015}\natexlab{}.
\newblock \showarticletitle{FloodGuard: A DoS Attack Prevention Extension in
  Software-Defined Networks}. In \bibinfo{booktitle}{\emph{2015 45th Annual
  IEEE/IFIP International Conference on Dependable Systems and Networks}}.
  \bibinfo{pages}{239--250}.
\newblock
\showISSN{1530-0889}
\urldef\tempurl%
\url{https://doi.org/10.1109/DSN.2015.27}
\showDOI{\tempurl}


\bibitem[\protect\citeauthoryear{Whaley}{Whaley}{1982}]%
        {whaley1982}
\bibfield{author}{\bibinfo{person}{Barton Whaley}.}
  \bibinfo{year}{1982}\natexlab{}.
\newblock \showarticletitle{Toward a general theory of deception}.
\newblock \bibinfo{journal}{\emph{The Journal of Strategic Studies}}
  \bibinfo{volume}{5}, \bibinfo{number}{1} (\bibinfo{year}{1982}),
  \bibinfo{pages}{178--192}.
\newblock


\bibitem[\protect\citeauthoryear{Ye}{Ye}{2018}]%
        {ye18}
\bibfield{author}{\bibinfo{person}{GenShen Ye}.}
  \bibinfo{year}{2018}\natexlab{}.
\newblock \bibinfo{title}{{75,000+ MikroTik Routers Are Forwarding Owners'
  Traffic to the Attackers, How is Yours?}}
\newblock \bibinfo{howpublished}{Netlab 360}.
\newblock
\newblock
\shownote{\url{https://blog.netlab.360.com/7500-mikrotik-routers-are-forwarding-owners-traffic-to-the-attackers-how-is-yours-en/}.}


\bibitem[\protect\citeauthoryear{Yoon, Park, Lee, Kang, Shin, and Zhang}{Yoon
  et~al\mbox{.}}{2015}]%
        {YOON:2015:SDN}
\bibfield{author}{\bibinfo{person}{Changhoon Yoon}, \bibinfo{person}{Taejune
  Park}, \bibinfo{person}{Seungsoo Lee}, \bibinfo{person}{Heedo Kang},
  \bibinfo{person}{Seungwon Shin}, {and} \bibinfo{person}{Zonghua Zhang}.}
  \bibinfo{year}{2015}\natexlab{}.
\newblock \showarticletitle{Enabling security functions with SDN: A feasibility
  study}.
\newblock \bibinfo{journal}{\emph{Computer Networks}}  \bibinfo{volume}{85}
  (\bibinfo{year}{2015}), \bibinfo{pages}{19 -- 35}.
\newblock
\showISSN{1389-1286}
\urldef\tempurl%
\url{https://doi.org/10.1016/j.comnet.2015.05.005}
\showDOI{\tempurl}


\bibitem[\protect\citeauthoryear{Yuill, Denning, and Feer}{Yuill
  et~al\mbox{.}}{2006}]%
        {deception:2006}
\bibfield{author}{\bibinfo{person}{Jim Yuill}, \bibinfo{person}{Dorothy
  Denning}, {and} \bibinfo{person}{Fred Feer}.}
  \bibinfo{year}{2006}\natexlab{}.
\newblock \bibinfo{title}{Using Deception to Hide Things from Hackers:
  Processes, Principles, and Techniques}.
\newblock \bibinfo{howpublished}{Journal of Information Warfare}.
\newblock


\bibitem[\protect\citeauthoryear{Zhang, Li, Xu, Bi, Gu, and Bai}{Zhang
  et~al\mbox{.}}{2018}]%
        {Zhang:reflection:2018}
\bibfield{author}{\bibinfo{person}{Menghao Zhang}, \bibinfo{person}{Guanyu Li},
  \bibinfo{person}{Lei Xu}, \bibinfo{person}{Jun Bi}, \bibinfo{person}{Guofei
  Gu}, {and} \bibinfo{person}{Jiasong Bai}.} \bibinfo{year}{2018}\natexlab{}.
\newblock \showarticletitle{Control Plane Reflection Attacks in {SDN}s: New
  Attacks and Countermeasures}. In \bibinfo{booktitle}{\emph{Research in
  Attacks, Intrusions, and Defenses}},
  \bibfield{editor}{\bibinfo{person}{Michael Bailey}, \bibinfo{person}{Thorsten
  Holz}, \bibinfo{person}{Manolis Stamatogiannakis}, {and}
  \bibinfo{person}{Sotiris Ioannidis}} (Eds.). \bibinfo{publisher}{Springer
  International Publishing}, \bibinfo{address}{Cham},
  \bibinfo{pages}{161--183}.
\newblock
\showISBNx{978-3-030-00470-5}


\bibitem[\protect\citeauthoryear{{Zhou}, {Wu}, {Yang}, {Wang}, {Yang}, {Lu},
  and {Cheng}}{{Zhou} et~al\mbox{.}}{2018}]%
        {SDN-RDC:2018}
\bibfield{author}{\bibinfo{person}{H. {Zhou}}, \bibinfo{person}{C. {Wu}},
  \bibinfo{person}{C. {Yang}}, \bibinfo{person}{P. {Wang}}, \bibinfo{person}{Q.
  {Yang}}, \bibinfo{person}{Z. {Lu}}, {and} \bibinfo{person}{Q. {Cheng}}.}
  \bibinfo{year}{2018}\natexlab{}.
\newblock \showarticletitle{SDN-RDCD: A Real-Time and Reliable Method for
  Detecting Compromised SDN Devices}.
\newblock \bibinfo{journal}{\emph{IEEE/ACM Transactions on Networking}}
  \bibinfo{volume}{26}, \bibinfo{number}{5} (\bibinfo{year}{2018}),
  \bibinfo{pages}{2048--2061}.
\newblock


\end{thebibliography}



\end{document}